\documentclass{emulateapj}
\usepackage{natbib}
\usepackage{color}

\shortauthors{T. Hosokawa et al.}
\shorttitle{Supermassive Supergiant Stars}

\newcommand{\msun}{M_{\odot}}
\newcommand{\rsun}{R_{\odot}}
\newcommand{\lsun}{L_{\odot}}
\newcommand{\msunyr}{M_\odot~{\rm yr}^{-1}}
\newcommand{\mdot}{\dot{M}_*}

\begin{document}

\title{Formation of Primordial Supermassive Stars \\
by Rapid Mass Accretion}
\author{Takashi Hosokawa\altaffilmark{1,2},
        Harold W. Yorke\altaffilmark{2},
        Kohei Inayoshi\altaffilmark{3}, \\
        Kazuyuki Omukai\altaffilmark{4,3},
        Naoki Yoshida\altaffilmark{1,5}}

\altaffiltext{1}{Department of Physics and 
Research Center for the Early Universe,
The University of Tokyo, Tokyo 113-0033, Japan;
takashi.hosokawa@phys.s.u-tokyo.ac.jp, hosokwtk@gmail.com}
\altaffiltext{2}{Jet Propulsion Laboratory, California Institute
of Technology, Pasadena CA 91109, USA}
\altaffiltext{3}{Department of Physics, Kyoto University, 
Kyoto 606-8502, Japan}
\altaffiltext{4}{Astronomical Institute, Tohoku University, Sendai
980-8578, Japan}
\altaffiltext{5}{Kavli Institute for the Physics and Mathematics
of the Universe, University of Tokyo, Kashiwa, Chiba
277-8583, Japan}

\begin{abstract}
Supermassive stars (SMSs) forming via very rapid mass accretion
($\mdot \gtrsim 0.1~\msunyr$) could 
be precursors of supermassive black holes observed beyond 
redshift of about 6.
Extending our previous work, we here study the evolution 
of primordial stars growing under such rapid mass accretion 
until the stellar mass reaches $10^{4 - 5}~\msun$.
Our stellar evolution calculations show that a star becomes 
supermassive while passing through the ``supergiant protostar'' stage, 
whereby the star has a very bloated envelope and a contracting inner core. 
The stellar radius increases monotonically with the stellar mass,
until $\simeq 100$~AU for $M_* \gtrsim 10^4~\msun$, 
after which the star begins to slowly contract.
Because of the large radius the effective temperature is always 
less than $10^4$~K during rapid accretion.  
The accreting material is thus almost completely 
transparent to the stellar radiation. 
Only for $M_* \gtrsim 10^5  M_\odot$ can stellar UV feedback
operate and disturb the mass accretion flow.
We also examine the pulsation stability of accreting
SMSs, showing that the pulsation-driven mass loss 
does not prevent stellar mass growth.
Observational signatures of bloated SMSs should be detectable 
with future observational facilities such as the James Webb 
Space Telescope. 
Our results predict that an inner core of the 
accreting SMS should suffer from the general relativistic 
instability soon after the stellar mass exceeds $10^5~\msun$. 
An extremely massive black hole should form after the collapse 
of the inner core. 
\end{abstract}


\section{Introduction}
\label{sec:intro}
\setcounter{footnote}{0}

Theoretical scenarios of forming supermassive black holes (SMBHs)
confront the observations of bright quasars beyond redshift 6, 
which require that SMBHs exceeding $10^9~\msun$ 
form in less than 1~Gyr after the Big Bang
\citep[e.g.,][]{Tr11,Mortlock11}.
The direct collapse of supermassive stars 
(SMSs; $M_* \gtrsim 10^5~\msun$)
could be a crucial first step leading to the rapid formation 
of SMBHs in the early universe.
If an SMS were to form, its inevitable collapse into a very massive black 
hole would be initiated by the general relativistic (GR) instability
\citep[e.g.,][]{Iben63,Chandra64}.
The $\sim$$10^5~\msun$ black holes thus generated could further grow
via subsequent mass accretion and mergers 
\citep[e.g.,][and references therein]{Haiman13}, and finally become
the SMBHs powering young bright quasars \citep[e.g.,][]{Johnson13c}.


The formation of an SMS should be understood in the context of
primordial star formation
\citep[e.g.,][]{BYHM09}. Recent theoretical work investigates 
how massive such primordial stars are, studying the late evolutionary 
stage whereby a protostar grows in mass via 
mass accretion from a surrounding envelope
\citep[e.g.,][]{OP03,MT08,Stacy10,HOYY11,
HYOY12,Stacy12,Susa13}.
Although the exact shape of their mass spectrum is still
under debate, recent studies show that primordial stellar 
masses should be widely distributed
from a few tens to hundreds of solar masses 
\citep[e.g.,][]{Susa13,Hirano13}. 
It is likely that the first stars are normally less massive 
than $10^3~\msun$. However, SMSs with $M_* \gg 10^3~\msun$ could 
nevertheless form through a non-standard mode of 
primordial star formation, in analogy to the present-day formation of
a massive star exceeding $100~\msun$, which although very rare, 
can form under non-standard
initial conditions which differ from those for low-mass 
($M_* \sim 1~\msun$) stars \citep[e.g.,][]{ZY07}.


A non-standard condition for forming the SMSs could
be realized when the formation of H$_2$ molecules, the primary coolant
in the primordial gas, is hindered in a massive dark halo 
\citep[e.g.,][]{BL03}.
A number of processes preventing H$_2$ formation have been proposed,
e.g., photodissociation by strong UV background
\citep[][]{Omukai01,OH02,WTA08,Shang10,WH11,IO11,Agarwal12}, 
and collisional dissociation in dense shocks \citep[][]{IO12}. 
Without H$_2$ molecules a primordial gas cloud collapses nearly 
isothermally at $T \simeq 8000$~K due to atomic hydrogen cooling.
Numerical models suggest that one or more supermassive stars 
may form in either spherical infall or disks 
\citep[e.g.,][]{BL03, RH09, Latif13, Latif13b}. 
After that, the stellar mass increases via very rapid mass accretion
at $\mdot \sim 0.1 - 1~\msunyr$ \citep[e.g.,][]{Latif13,Choi13}.
The star can become supermassive within $10^6$~yr with
such rapid mass accretion.
Note this process is operative only in an almost pristine gas 
with metallicity $\la 10^{-5} Z_{\sun}$, since fragmentation induced 
by dust cooling would prohibit the monolithic collapse of
massive clouds at higher metalicities \citep{OSH08}.


The evolution of accreting SMSs is critical for understanding 
their formation process. For instance, the stellar effective temperature 
controls energies of photons emitted by an SMS, whose luminosity 
could exceed $10^9~L_\odot$ at $M_* \sim 10^5~\msun$. 
If the SMS emits a copious amount of UV photons, 
an H{\sc ii} region should form around the SMS.
\cite{Johnson12} show that, under spherical symmetry, 
the resulting UV feedback halts the mass accretion only 
for $M_* \gtrsim 10^6~\msun$. 
In disk geometries, however, a bipolar H{\sc ii} region should emerge
and strong UV feedback could limit mass accretion much earlier, 
as expected in the standard cases of primordial star formation 
\citep[e.g.,][]{MT08,HOYY11,Stacy12,Susa13}. 
The final mass of the SMS should depend on the details of 
its evolution before the rapid mass accretion ceases. 


The evolution of SMSs with various metallicities
has been modeled in detail
\citep[e.g.,][]{Osaki66,Unno71,AF72a,AF72b,Fricke73,Fuller86}.
However, they do not consider the formation stage with rapid 
mass accretion, which could significantly change 
the stellar structure \citep[e.g.,][]{Begelman10,HOY12,Sch13}.
The ultimate fate of an SMS, i.e., whether a star collapses
to form a BH \citep[e.g.,][]{Shibata02, Montero12, Reisswig13}, 
explodes by thermo-nuclear burning 
\citep[e.g.,][]{Whalen12,Johnson13b,Whalen13b,Whalen13}, 
or evolves into a ``quasi-star'' \citep{Begelman08}, 
also depends on the mass accretion history during the formation stage.


\citet[][hereafter HOY12]{HOY12} study stellar evolution with 
very high accretion rates $\mdot \gtrsim 10^{-2}~\msunyr$,
employing numerical codes developed 
in our previous work \citep[e.g.,][]{OP03,HO09,HYO10}.
Our calculations show that the evolution at such
high accretion rates qualitatively differs from 
that with lower accretion $\mdot \lesssim 10^{-2}~\msunyr$,
normally expected for primordial star formation. At high accretion rates
the stellar radius is found to increase monotonically with mass 
\begin{equation}
R_* \simeq 2.6 \times 10^3~R_\odot 
\left( \frac{M_*}{100~M_\odot} \right)^{1/2} 
\label{eq:rst_analytic}
\end{equation}
for $M_* \gtrsim 100~\msun$, independent of the 
accretion rate. The stellar effective temperature is almost 
constant at $\simeq 5000$~K during this stage, and the resulting 
stellar UV luminosity is very low. 
However, HOY12 could only focus on the early evolution up to 
stellar masses of $10^3~\msun$, 
because their iteration procedure for constructing 
stellar models converged very slowly at higher masses.
Here, using a different stellar evolution code \citep[][]{YB08},
we extend the previous work of HOY12 and discuss
the evolution up to stellar masses $10^{4 - 5}~\msun$.
Moreover, we consider the pulsation stability of accreting
SMSs extending \citet[][hereafter IHO13]{IHO13}, whose analyses are 
limited to $M_* \leq 10^3~\msun$. We examine whether the resulting mass 
loss could limit stellar growth via rapid mass accretion.


This paper is organized as follows. First, we briefly describe the 
numerical methods adopted and cases considered in Section~\ref{sec:model}.
The results are presented in Section~\ref{sec:results} and
compared with our 
previous work in HOY12. We provide discussions and concluding 
remarks in Sections~\ref{sec:discussions} and \ref{sec:conclusions}.

\section{Numerical Modeling of Accreting Stars}
\label{sec:model}

\subsection{Evolutionary Calculations}
\label{ssec:num}


We study the evolution of rapidly accreting stars
by numerically solving their interior structure.
The governing equations are the usual stellar structure equations
including effects of the mass accretion \citep[e.g.,][]{Kip77,SST80}.
In this paper, we employ the numerical code developed by
\citet{YB08}, where the Henyey method \citep{Henyey64} is employed 
to solve the stellar interior structure \citep[also see][]{Bod07}.
The outer boundary conditions for the interior are
provided by solving the structure of a gray atmosphere.
To do this, we integrate inward the equations of hydrostatic 
equilibrium and radiative or convective transfer 
over the atmosphere with a Runge-Kutta procedure.
The stellar effective temperature $T_{\rm eff}$ is defined
at the optical depth $\tau = 2/3$ measured from the exterior.
The atmosphere integration extends to a connecting point
at a temperature 
$T_{\rm atm} \simeq {\rm several} \times T_{\rm eff}$, 
which varies slowly with time.
We presume that the thermal state of the accreting material
is the same as in the outer atmosphere (see below). 
The material added to the atmosphere is ultimately incorporated 
into the interior by means of an automatic rezoning procedure.


The current numerical code differs from that used in HOY12
\citep[also see][]{HO09}, 
where the same governing equations were solved by a
shooting method instead of the Henyey method used here.
Inside an SMS radiation dominates the total pressure,  
but a small contribution by the gas is crucial in determining its 
structure such as the radius. Here we adopt the gas pressure as one  
of our independent Henyey variables in place of the total pressure. 
This enables us to accurately solve the structure of a SMS.
The outer boundary conditions are also different. 
Most of the cases examined in HOY12 assume spherical
accretion onto a star \citep[e.g.,][]{SST80,HO09}, whereby
steady accretion flow hits the stellar surface forming
an accretion shock front.
In this case, we also consider the structure of the 
steady accretion flow falling onto the star.
We calculate the radial structure of both the outer accretion flow
and stellar interior for satisfying the shock jump conditions.


HOY12 also study a few cases with a different outer boundary, 
the photospheric boundary conditions: 
\begin{equation}
P_{\rm surf} = \frac23 \frac{1}{\kappa_{\rm surf}} 
\frac{G M_*}{R_*^2},
\label{eq:photo_b1}
\end{equation}
and 
\begin{equation}
L_* = 4 \pi R_*^2 \sigma T_{\rm surf}^4 
\label{eq:photo_b2}
\end{equation}
\citep[e.g.,][]{PS92,HYO10}, 
where the suffix ``surf'' represents quantities at the stellar 
surface and $\sigma$ is the Stefan-Boltzmann constant.
This boundary is rather similar to that adopted in the current code, 
except that the structure of the atmosphere is not solved. 
These outer boundary conditions assume that the accreting material
is added to the star through a geometrically thin disk, whereby most of 
the stellar surface can still freely radiate (no backwarming).


Although one can see that the photospheric
boundary is realistic for the assumed disk accretion, 
it is important that the outer boundary condition correctly determines 
the specific entropy of accreting material
\citep[e.g.,][]{HCK97,HOK11}. 
With the photospheric boundary, the accreting material
has the smallest amount of entropy (so-called {\it cold accretion}), 
which is the value at the stellar surface.
In this case, accreting gas would slowly approach 
the star through the disk, radiating its heat 
away before reaching the stellar surface.
The accreting gas gently lands on the stellar
surface with the same entropy as in the outer photosphere. 
With the spherically symmetric shock boundary condition, 
on the other hand, the accreting material has a much higher 
entropy than for the cold accretion case, 
carrying a larger fraction of the entropy 
generated in the shock front into the stellar interior.
However, even for the case of highly non-spherical accretion through
a circumstellar disk, the very rapid mass accretion considered
here means that a non-negligible amount of entropy should be 
advected into the star.
For this reason, HOY12 adopt the shock boundary condition 
for the fiducial cases.


Detailed modeling of the thermal properties of 
the accretion flow is beyond the scope of
spherically symmetric stellar evolution calculations. 
Realizing that a fraction of the gravitational energy 
released by the accreting gas will be deposited into the stellar
interior \citep[e.g.,][]{Siess96,Siess97}, 
we add an additional luminosity
\begin{equation}
L_{*,{\rm acc}} \equiv \eta L_{\rm acc}
= \eta \frac{G M_* \dot{M}_*}{R_*}
\label{eq:eta}
\end{equation}
to the bottom of the atmosphere.
The fraction of the deposited accretion luminosity $\eta$ is 
treated as a free parameter with a value $0 \leq \eta \leq 1$. 
The limit of cold accretion corresponds to $\eta = 0$.
More detailed modeling would be possible if we were to consider how
the deposited energy is spatially distributed 
in the stellar interior \citep[e.g.,][]{Siess96}, i.e. 
not only at the bottom of the atmosphere.
However, we find that the evolution does not depend on $\eta$
for $M_* \gtrsim 100~\msun$.

\subsection{Cases Considered}
\label{ssec:cases}

Table 1 summarizes the cases considered in this paper.
We follow the stellar evolution with rapid mass accretion at
various constant rates of 0.01, 0.1, 0.3, 
and $1.0~\msunyr$ beginning with a $2~\msun$ initial model.
Cases with different values of $\eta$ 
are also examined for each accretion rate.
Although the current numerical code enables following
the evolution far beyond that considered by HOY12, 
we ultimately encounter numerical convergence difficulties. 
Here we present the evolution for $\simeq 10^5$ years in each case.  
The stellar mass reaches $10^5~\msun$ 
with the accretion rate $\mdot = 1.0~\msunyr$ 
(cases a1, see table 1).

The cases with the lowest accretion rate of $0.01~\msunyr$ 
(cases a0.01) are relevant to
the normal mode of the primordial star formation, 
where H$_2$ molecular cooling operates. 
We also study these cases as our previous calculations predict 
that, with such moderately high accretion rates,
the stellar radius shows a peculiar behavior, 
oscillating with increasing stellar mass
\citep[e.g.,][HOY12]{OP01,OP03}.
Below, we show that our current numerical code also provides
a similar evolution despite a number of 
differences, e.g., the outer boundary conditions.

\begin{table}[t]
\label{tb:md}
\begin{center}
Table 1. Cases Considered \\[3mm]
\begin{tabular}{cccc}
\hline
\hline
Cases & $\mdot$ ($\msunyr$) & $\eta$ & $M_{*,f}(\msun)$ \\
\hline
a1-e0      & $1.0$ &  $0.0$  & $10^5$ \\
a1-e0.01   & $1.0$ &  $0.01$ & $10^5$ \\
a1-e0.1    & $1.0$ &  $0.1$  & $10^5$ \\
a1-e1      & $1.0$ &  $1.0$  & $10^5$ \\
a0.3-e0.1  & $0.3$ &  $0.1$  & $6 \times 10^4$ \\
a0.1-e0    & $0.1$ &  $0.0$  & $2 \times 10^4$ \\
a0.1-e0.01 & $0.1$ &  $0.01$ & $2 \times 10^4$ \\
a0.1-e0.1  & $0.1$ &  $0.1$  & $2 \times 10^4$ \\
a0.1-e1    & $0.1$ &  $1.0$  & $2 \times 10^4$ \\
a0.01-e0.1 & $0.01$ &  $0.1$ & $2 \times 10^3$ \\
a0.01-e0   & $0.01$ &  $0.0$ & $2 \times 10^3$  \\
\hline
\end{tabular}
\noindent
\end{center}
Col.\ 2: mass accretion rate, 
Col.\ 3: fraction of the accretion luminosity deposited 
in the stellar interior (see eq.~\ref{eq:eta}), 
Col.\ 4: final stellar mass.
\end{table}

\section{Results}
\label{sec:results}

\subsection{Evolution of Rapidly Accreting Supermassive Stars}
\label{ssec:evolution}

\begin{figure}
  \begin{center}
\epsscale{1.0}
\plotone{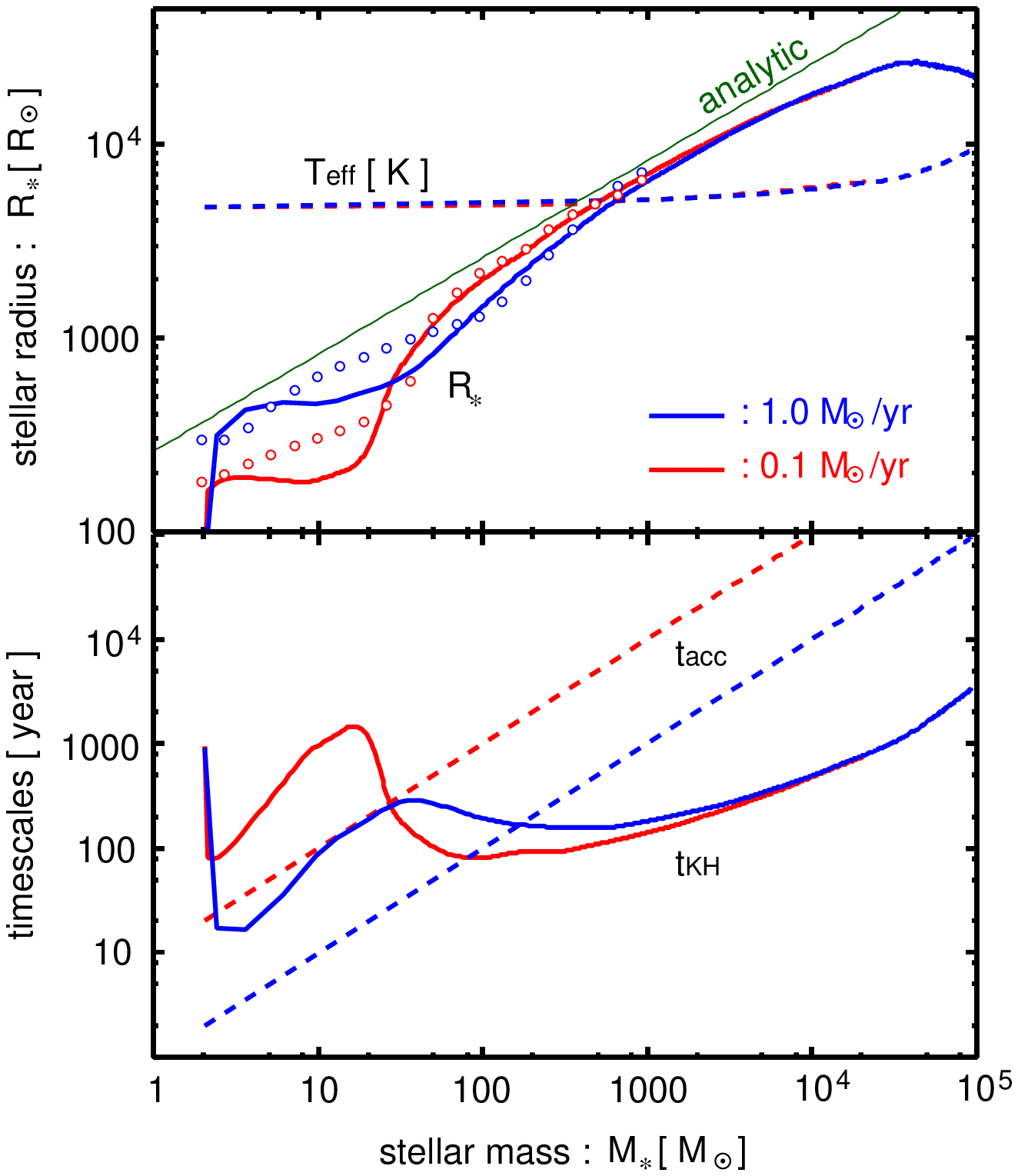}
\caption{Protostellar evolution over $\sim 10^5$ years with 
accretion rates $\mdot = 0.1~\msunyr$ (case a0.1-e0.1, red lines) 
and $1.0~\msunyr$ (case a1-e0.1, blue lines), 
and $\eta=0.1$ (see text). 
{\it Upper panel:} evolution of the stellar radius (solid lines).
The open circles represent our previous results 
taken from \citet{HOY12} 
for $M_* \lesssim 10^3~M_\odot$ with $\mdot = 0.1~\msunyr$ (red) 
and $1.0~\msunyr$ (blue).
The thin green line shows the analytic mass-radius relation
given by equation (\ref{eq:rst_analytic}).
The evolution of the stellar effective temperature is also
overlaid on the above plots (dashed lines),
using the same scale as for the stellar radius. 
{\it Lower panel:} evolution of the accretion timescale 
(dashed lines) and Kelvin-Helmholz timescale (solid lines).
}
\label{fig:mr_tscl}
  \end{center}
\end{figure}

\subsubsection{Outline of the Evolution}
\label{ssec:outline}


Figure~\ref{fig:mr_tscl} presents the overall stellar evolution with 
the accretion rates $\mdot = 0.1~\msunyr$ (case a0.1-e0.1) and 
$1.0~\msunyr$ (case a1-e0.1), where the deposited energy fraction 
$\eta$ is set at 0.1. 
We see that, in both cases, the early evolution of the stellar 
radius agrees well with our previous results for 
$M_* \lesssim 10^3~\msun$ (HOY12).
The small deviations for $M_* \lesssim 100~\msun$ are mostly 
due to the different outer boundary conditions as explained in 
Sec.~\ref{ssec:num} (also see Sec.~\ref{ssec:eta_dep} below).
Our current results show that the star continues to expand 
for $M_* \gtrsim 10^3~\msun$.
The stellar radius increases monotonically with mass
according to relation (\ref{eq:rst_analytic}) for
$100~\msun \lesssim M_* \lesssim 10^4~\msun$, 
and finally begins to decrease for $M_* \gtrsim 3 \times 10^4~\msun$.
Nonetheless, the stellar radius remains very large
at $R_* \simeq 2 \times 10^4~\rsun \simeq 100$~AU.
With the accretion rate of $\mdot = 1.0~\msunyr$, we obtain
an accreting $10^5~\msun$ star which is still significantly bloated.


To understand why the radius begins to decrease for 
$M_* \gtrsim 10^4~\msun$, it is helpful to recall how 
equation (\ref{eq:rst_analytic}) is derived (also see HOY12). 
In general, the luminosity of very massive stars is close 
to the Eddington value, 
\begin{equation}
L_* = 4 \pi R_*^2 \sigma T_{\rm eff}^4 
\simeq L_{\rm Edd} \propto M_*.
\label{eq:rst_org}
\end{equation}
Because of the very strong temperature-dependence of gas opacity with
H$^-$ bound-free absorption, which dominates near the stellar surface,
the effective temperature is locked at several $\times$ $10^3$~K 
\citep[e.g.,][]{Hayashi61}. Equation (\ref{eq:rst_analytic}) is derived
by substituting $T_{\rm eff} = 5000$~K into (\ref{eq:rst_org}).
The upper panel of Figure~\ref{fig:mr_tscl} also shows that 
the effective temperature is certainly locked at 
$\simeq 5000$~K for $M_* \lesssim 10^4~\msun$. 
However, we see that the effective temperature gradually 
increases for $M_* \gtrsim 10^4~\msun$, 
when the stellar expansion ceases. 
The deviation of the stellar expansion from relation 
(\ref{eq:rst_analytic}) is due to the rise of
effective temperature, which is separately considered
in Sec.~\ref{ssec:surf} below.


In general, the evolution of an accreting star can be understood
by comparing the accretion time
\begin{equation}
t_{\rm acc} \equiv \frac{M_*}{\mdot} , 
\end{equation}
the timescale for stellar mass growth,
to the Kelvin-Helmholtz (KH) time
\begin{equation}
t_{\rm KH} \equiv \frac{G M_*^2}{R_* L_*} ,
\end{equation}
over which the star radiates away its gravitational energy
\citep[e.g.,][]{SPS86}.
The lower panel of Figure~\ref{fig:mr_tscl} shows that
the balance between these two timescales is inverted 
from $t_{\rm KH} > t_{\rm acc}$ to $t_{\rm KH} < t_{\rm acc}$ 
during the evolution, e.g., 
at $M_* \simeq 200~\msun$ for $1.0~\msunyr$.
As explained in the literature \citep[e.g.,][]{OP03,HO09},
this is because the KH timescale decreases as 
the stellar luminosity $L_*$ rapidly
increases with stellar mass. 
The same timescale inversion is also shown in HOY12
(see their Fig.~3).


We see that the accreting stars have a very short KH timescale 
compared to the accretion timescale for 
$M_* \ga {\rm a\ few} \times 10^2~\msun$.
This means that the star efficiently loses its energy via
radiation, which usually leads to contraction
(so-called KH contraction). 
However, our results show that the stellar radius continues to increase until
$M_* \gtrsim 3 \times 10^4~\msun$.
This suggests that the star has a very inhomogeneous structure, e.g., 
most of the stellar interior contracts while radiating energy away,
whereas a surface layer expands as it receives part of the outward 
heat flux. HOY12 show this core-envelope structure of 
rapidly accreting stars for $M_* \leq 10^3~\msun$.
In Sec.~\ref{ssec:intr} below, we show that the same stellar structure 
is maintained even after the star becomes supermassive, 
$M_* \gtrsim 10^4~\msun$.

\subsubsection{Stellar Interior Structure }
\label{ssec:intr}

\begin{figure}
  \begin{center}
\epsscale{1.1}
\plotone{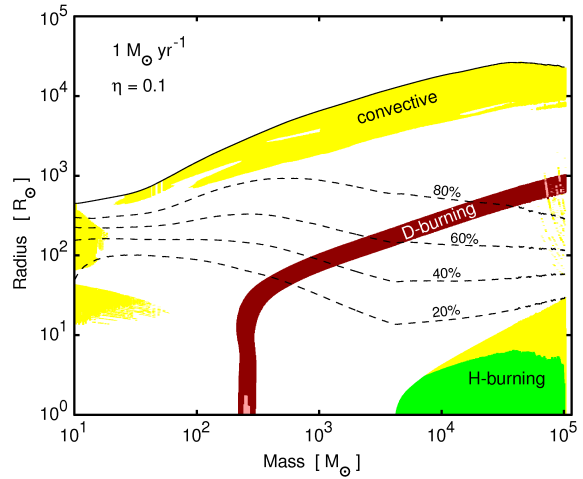}
\caption{Evolution of the stellar interior structure 
until the accreted stellar mass reaches $10^5~\msun$
for the accretion rate $\mdot = 1~\msunyr$ and 
energy fraction $\eta = 0.1$ (case a1-e0.1).
The black solid and dashed lines represent the radial positions 
of the stellar surface and mass
coordinates of $80\%$, $60\%$, $40\%$, and $20\%$ of the total
stellar mass in descending order.
The white and yellow areas represent radiative and convective 
zones without nuclear fusion, respectively. 
The brown stripe denotes the radiative layer where deuterium
burning occurs, i.e., the energy production rate exceeds 
the mass-averaged steady rate with mass accretion
$L_{\rm D,st}/M_*$ (see equation~\ref{eq:ldst}).
The green area represents a convective core where hydrogen 
burning occurs, i.e., the hydrogen depletion timescale is 
shorter than the lifetime of the main-sequence star with the same mass. 
Pink zones depict convective deuterium burning.}
\label{fig:intr_1em0En0.1}
  \end{center}
\end{figure}
\begin{figure}
  \begin{center}
\epsscale{1.0}
\plotone{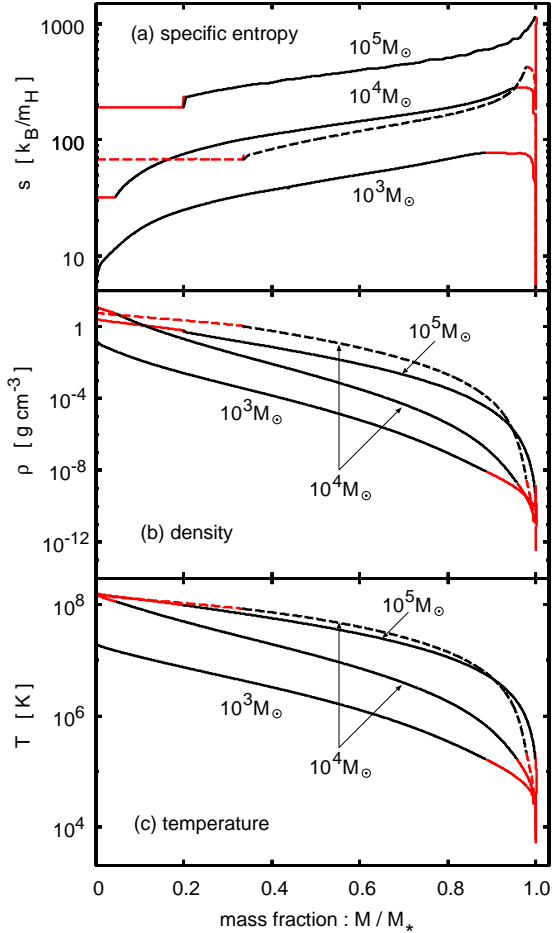}
\caption{Snapshots of the thermal and density structure in the stellar interior.
The solid and dashed lines represent the two accretion
rates $1.0~\msunyr$ and $0.1~\msunyr$ with the same input
energy fraction $\eta = 0.1$ (cases a1-e0.1 and a0.1-e0.1).
The panels ({\it a}), ({\it b}), and ({\it c}) show the radial 
distributions of the specific entropy, density, and temperature as 
functions of the normalized mass coordinate 
$M / M_*$. For the accretion rate $1.0~\msunyr$ the structure after
$10^3~\msun$, $10^4~\msun$, and $10^5~\msun$ have accreted 
is displayed. 
Profiles at the lower accretion rate are shown
only for the $10^4~\msun$ star.
The red parts denote the convective layers.}
\label{fig:rprof}
  \end{center}
\end{figure}
\begin{figure}
  \begin{center}
\epsscale{1.0}
\plotone{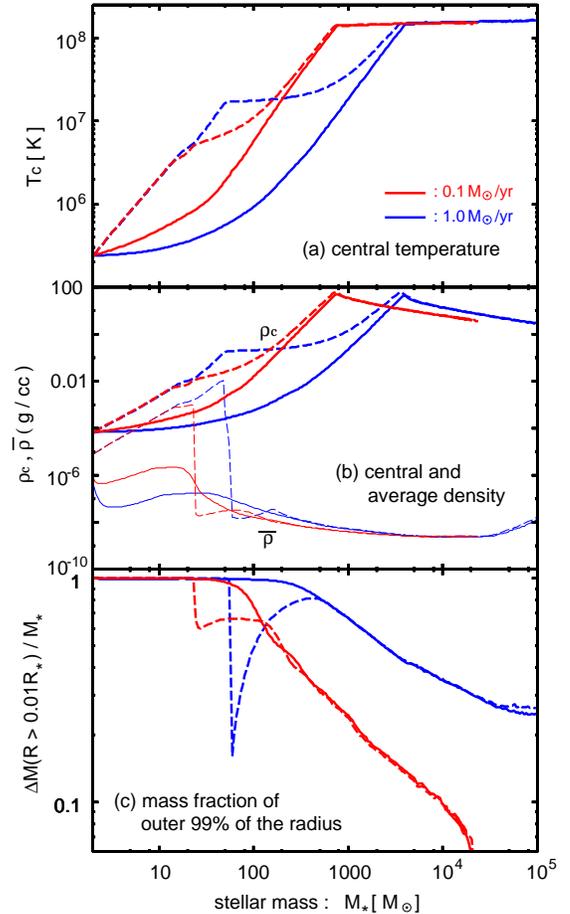}
\caption{
Variations of the stellar interior structure with different
accretion rates $\mdot $ and input energy fractions $\eta$.
The panels ({\it a}), ({\it b}), and ({\it c}) show the
evolution of the central temperature $T_{\rm c}$, 
the central density $\rho_{\rm c}$,
the average density $\bar{\rho}$, 
and the mass fraction of an outer layer which covers 99~\% of the 
stellar radius $\Delta M (R > 0.01~R_*)/M_*$.
The blue and red colors denote the different accretion rates
$1.0~\msunyr$ and $0.1~\msunyr$, respectively.
The solid and dashed lines represent deposited energy 
fractions of $\eta = 0.1$ (cases a1-e0.1 and a0.1-e0)
and $0.0$ (cases a0.1-e0.1 and a0.1-e0).
}
\label{fig:int_var}
  \end{center}
\end{figure}

We here investigate the evolution of the interior structure
of accreting supermassive stars. 
Figure~\ref{fig:intr_1em0En0.1} shows the evolution 
with the accretion rate $\mdot = 1.0~\msunyr$ and input energy 
fraction $\eta = 0.1$ (case a1-e0.1). 
As explained in Sec.~\ref{ssec:outline}, the KH timescale falls
below the accretion time at $M_* \simeq 200~\msun$ 
in this case.


Although deuterium burning begins soon after the timescale 
inversion, this hardly influences the subsequent evolution 
(also see HOY12). At this stage deuterium burns at nearly a steady 
rate, because it is replenished by accretion as quickly as consumed 
by burning, 
\begin{equation}
L_{\rm D,st} \equiv \mdot \delta_{\rm D}
= 1.5 \times 10^6 \lsun 
\left( \frac{\mdot}{1~\msunyr}  \right)
\left( \frac{[{\rm D}/{\rm H}]}{2.5 \times 10^{-5}}  \right) ,
\label{eq:ldst}
\end{equation}
where $\delta_{\rm D}$ is the energy produced via deuterium
burning per unit mass. The energy released by deuterium burning 
is negligible in comparison to the local luminosity, 
approximately given by the Eddington value.
The heat generated by nuclear fusion is rapidly transported 
via radiation without causing convection.
This is not the case with extremely cold mass accretion,
$\eta = 0$ (see below).


Figure~\ref{fig:intr_1em0En0.1} shows that for 
$M_* \gtrsim 200~\msun$ the star has an inhomogeneous 
mass distribution; a surface layer which contains only a tiny 
fraction of the total stellar mass significantly inflates.
Moreover, we see that the star becomes increasingly
centrally condensed as the stellar mass increases.
This is because most of the stellar interior is contracting
in this stage as explained in Sec.~\ref{ssec:outline}.
The gravitational energy released by the contracting
interior is radiatively transported outward.
However, part of the outward flux is absorbed
in a surface layer 
where the opacity is higher than in the interior.
As a result, the surface layer has a high specific entropy
and largely inflates.
Figure~\ref{fig:intr_1em0En0.1} shows that the energy is 
partly transported by convection in this semi-opaque surface layer.
Figure~\ref{fig:rprof}-(a) presents the characteristic profiles
of the specific entropy in the stellar interior, 
showing that the entropy increases outward  
to reach a maximum at the bottom of the surface convective layer.


Figure~\ref{fig:int_var}-(a) shows that, as in the ordinary KH 
contraction stage, the stellar central temperature increases 
with mass. 
Hydrogen burning begins at $M_* \simeq 4 \times 10^3~\msun$, 
after which the central temperature is fixed at 
$\simeq 1.5 \times 10^8$~K by the thermostat effect 
due to the strong temperature-dependence of energy production
through the CNO cycle.
Although initially at zero metallicity a small amount of carbon 
is produced in the center through the 3-$\alpha$ process during KH 
contraction, which allows efficient hydrogen nuclear burning via 
the CNO cycle.
Figures~\ref{fig:intr_1em0En0.1} and \ref{fig:int_var} 
show that a convective hydrogen-burning core grows with 
increasing stellar mass.
The ignition of hydrogen burning halts the KH contraction 
within and near the convective core. 
In fact, Figure~\ref{fig:int_var}-(b) shows that the ratio of the
central density to the average density decreases with 
increasing mass after hydrogen burning begins.
However, most of the stellar interior is still contracting
as can be inferred from Figure~\ref{fig:intr_1em0En0.1}.
As the mass increases from 
$10^4~\msun$ to $10^5~\msun$ the radii of 80\% and
60\% of the total mass continue to decrease and the radii of 40\%
and 20\% increase only slightly, but much less than the
growth in mass. Finally, at $M_* \simeq 10^5~\msun$ the outer 
99~\% of the star in radius contains only 25~\% of its total mass 
(see Figure~\ref{fig:int_var}-(c)).
Figure~\ref{fig:rprof}-(b) and (c) show that the average
interior density and temperature still rise
with increasing stellar mass as a result of the on-going 
contraction for $M_* \gtrsim 10^4~\msun$.
The gravitational energy released by the contracting
envelope compensates most of the stellar surface
luminosity, even after the ignition of hydrogen burning.
Most of the heat generated by nuclear fusion is
absorbed in the convective core, whose entropy is consequently 
elevated as shown in Figure~\ref{fig:int_var}-(a). 


Finally, we consider how the above evolution changes with 
a lower accretion rate $\mdot = 0.1~\msunyr$.
As can be inferred from Figure~\ref{fig:mr_tscl}, 
the evolution is qualitatively the same as in the previously discussed
cases with $\mdot = 1.0~\msunyr$. There is a quantitative
difference, however. At the lower accretion rate
the accretion timescale is correspondingly much longer than
the KH timescale for $M_* \gtrsim 100~\msun$.
The star has a more centrally-concentrated 
mass distribution for a given stellar mass, because it has had 
more time to contract and has radiated more energy $\int L_*\, dt$,
which has to be covered by gravitational contraction.
Figure~\ref{fig:int_var}-(a) shows that, as a result, the
central temperature is higher at comparable stellar 
masses, so that hydrogen burning begins before the stellar 
mass reaches $10^3~\msun$.
The outer envelope of the star still continues to contract
after the ignition of hydrogen as previously described.
Figure~\ref{fig:rprof}-(b) and (c) show that, 
for a $10^4~\msun$ star, the average interior temperature and 
density for $\mdot = 0.1~\msunyr$ are much higher than those
with $\mdot = 1.0~\msunyr$.
It is interesting that, regardless of the different interior structure, 
the evolution of the stellar radius, while obeying relation 
(\ref{eq:rst_analytic}), is almost independent of 
the accretion rate.

\subsubsection{Stellar Surface Structure and Termination of the Expansion}
\label{ssec:surf}

\begin{figure}
  \begin{center}
\epsscale{1.0}
\plotone{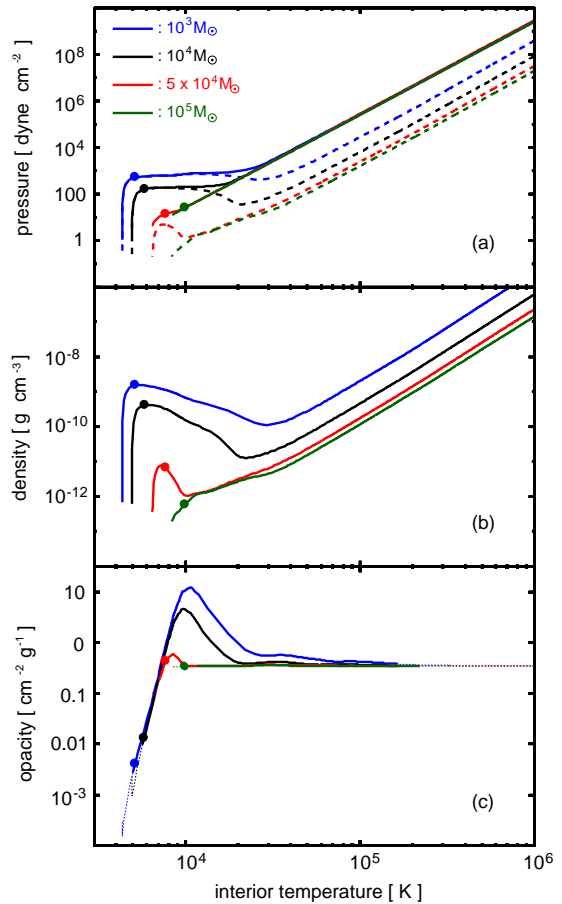}
\caption{Structure near the stellar surface
for the accretion rate $\mdot = 1~\msunyr$ and 
input energy fraction $\eta = 0.1$ (case a1-e0.1). 
The panels (a), (b), and (c) show the radial distributions of
pressure, density and opacity as functions of the 
interior temperature. 
Higher interior temperatures correspond to greater depths into the star.
The different colors denote the different 
epochs when the stellar mass is $10^3~\msun$ (blue), 
$10^4~\msun$ (black), $5 \times 10^4~\msun$ (red), 
and $10^5~\msun$ (green).
The filled circles on the lines mark the positions 
of the photosphere. In panel (a), the solid and dashed lines 
represent the total pressure and gas pressure, respectively. 
In panel (c), the solid and thin dotted lines denote
the convective and radiative layers, respectively.
For $T \gtrsim 10^5$~K the opacity converges
to the constant value from Thomson scattering.\
}
\label{fig:surface}
  \end{center}
\end{figure}

Our results show that rapidly accreting SMSs have a large radius
$\sim 100$~AU, even after the stellar mass exceeds $10^4~\msun$.
However, the stellar expansion gradually
slows down, no longer following the analytic mass-radius relation 
(\ref{eq:rst_analytic}). Figure~\ref{fig:mr_tscl} shows that, especially 
with $\mdot = 1.0~\msunyr$, the stellar radius finally turns around 
and decreases slightly for $M_* \gtrsim 3 \times 10^4~\msun$.
Here we consider why the stellar expansion finally ceases.


As explained in Sec.~\ref{ssec:intr}, the stellar structure is not 
homogeneous during expansion. 
Only a surface layer which has a tiny fraction of the total 
stellar mass inflates. Whereas most of the stellar 
interior contracts and radiates the energy away, the outer layer 
absorbs a part of the outward flux and gains a relatively
high specific entropy, resulting in the surface-layer bloating.  
The physical state of the outermost part of a star is thus key
to understand the termination of the expansion.
Figure~\ref{fig:surface} presents several snapshots of the
outermost radial profiles of pressure, density, and opacity 
as functions of the interior temperature $T$
for case a1-e0.1 ($\mdot = 1.0~\msunyr$ and $\eta = 0.1$).
Note that higher interior temperatures represent
deeper stellar layers, because the interior temperature 
increases monotonically toward the center.
For $M_* \lesssim 10^4~\msun$ while the stellar
radius is still increasing with mass according to
(\ref{eq:rst_analytic}), the star has a characteristic 
feature for $T \lesssim 3 \times 10^4$~K.
Panel (a) shows that the gas pressure dominates over the 
radiation pressure there, shaping the almost flat profile
within the photosphere. We also see that the density accordingly 
{\em increases} outward (panel-b), i.e. it is not monotonic. 
This structure is called a density inversion.
Such density inversions occasionally occur in the outermost
layers where the radiative flux locally exceeds 
the Eddington value.
The hydrostatic balance is achieved with the inward gas pressure 
gradient (around $T \simeq 2 \times 10^4$~K in panel-a),
which helps gravity balance the outward radiation force.
The gas opacity has a peak at $T \simeq 10^4$~K because of 
the bound-free absorption of H and H$^-$ \citep[e.g.,][]{Kip}. 
In the outer layers where $T \lesssim 10^4$~K, the opacity drops 
sharply with decreasing temperature, a general feature of 
H$^-$ absorption, because at lower temperatures there are fewer 
free electrons available for forming H$^-$ ions.


However, this characteristic feature at the surface
changes when the stellar mass exceeds $M_* > 10^4~\msun$.
Radiation pressure starts to dominate the total 
pressure everywhere in the interior, and the density inversion 
almost disappears by the epoch of $M_* \simeq 10^5~\msun$. 
This behavior is due to the density-dependence of gas opacity.
Figure~\ref{fig:surface}-(c) shows that, at the low density
$\rho \lesssim 10^{-11}~{\rm g\, cm}^{-3}$, opacity in the surface 
layers drops to the Thomson scattering value as H$^-$ absorption 
becomes inefficient \citep[also see, e.g.,][]{MD05}.
This is inevitable because collisional events between 
neutral hydrogen atoms and free electrons are required to 
form H$^-$ ions and the H$^-$ opacity is therefore proportional
to $n_{\rm H} n_{\rm e}$.

Recall that equation (\ref{eq:rst_analytic}) is derived assuming
that the effective temperature is locked at $T_{\rm eff} \simeq 5000$~K 
because of the strong temperature-dependence of H$^-$ opacity.
At the point when H$^-$ ions no longer dominate the opacity in
surface layers, these surface layers no longer 
efficiently absorb the heat flux coming from the contracting interior. 
We thus predict that the stellar radius  will continue to decrease
after the stellar mass exceeds $10^5~\msun$, 
in contrast to analytical considerations by \citet{Sch13},
who do not consider the detailed 
structure in the outermost part of the star where H$^-$ opacity 
is important. We stress that the density-dependence of H$^-$ opacity 
is a key for understanding the termination of the stellar expansion.

\subsubsection{Evolution with Different Input Energy Fractions $\eta$}
\label{ssec:eta_dep}

\begin{figure}
  \begin{center}
\epsscale{1.0}
\plotone{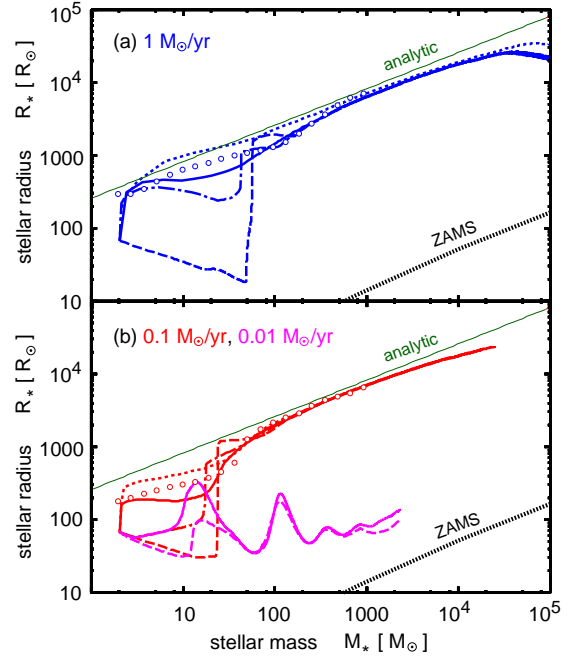}
\caption{Evolution of the stellar radius with 
different fractions of the accretion luminosity deposited
in the stellar interior, $\eta$.
The upper and lower panels show the evolution with the
accretion rates $\mdot = 1.0~\msunyr$, $0.1~\msunyr$, and $0.01~\msunyr$.
The dotted, solid, dot-dashed, and dashed lines represent 
$\eta = 1.0$ (cases a1-e1 and a0.1-e1), 
$0.1$ (a1-e0.1, a0.1-e0.1, and a0.01-e0.1), 
$0.01$ (a1-e0.01 and a0.1-e0.01), 
and $0.0$ (a1-e0, a0.1-e0, and a0.01-e0).
In each panel, the thin green line shows the analytic mass-radius relation
given by equation (\ref{eq:rst_analytic}). 
The open circles denote our previous results taken from
\citet{HOY12} as in Figure~\ref{fig:mr_tscl}.
The black dotted line represents the mass-radius relation of
ZAMS stars.}
\label{fig:mr_fe}
  \end{center}
\end{figure}
\begin{figure}
  \begin{center}
\epsscale{1.1}
\plotone{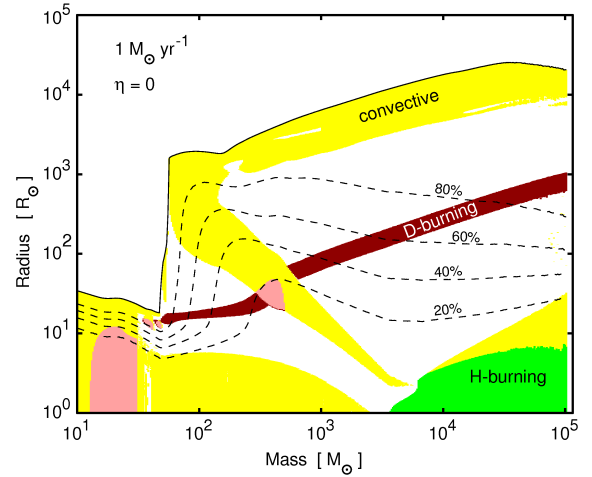}
\caption{Same as Figure~\ref{fig:intr_1em0En0.1} but for the
zero energy input $\eta = 0$ (case a1-e0). }
\label{fig:intr_1em0En0}
  \end{center}
\end{figure}

As described in Sec.~\ref{ssec:num}, we parametrize the 
gravitational energy deposited into the stellar interior with
the free parameter $\eta$.
Here, we investigate how the stellar evolution varies with
different values of $\eta$.
We naively expect that, at a fixed stellar mass,  
the stellar radius would be larger with higher $\eta$, 
as the specific entropy of accreting materials would be enhanced.
 Figure~\ref{fig:mr_fe} shows this is true only for early
evolutionary stages when $M_* \lesssim 200~\msun\ (30~\msun)$ 
for the accretion rate 
$\mdot = 1.0~\msunyr$\ ($0.1~\msunyr$, respectively), 
which corresponds to the epoch when the 
accretion timescale is shorter than the KH timescale
(see Sec.~\ref{ssec:outline}).
With this timescale imbalance, gas in the stellar interior retains
the entropy originally obtained when accreted
\citep{SPS86}, which in our case is determined by $\eta$.
Differences of $\eta$ thus influence the stellar structure
during this stage.
After the timescale inversion to $t_{\rm acc} > t_{\rm KH}$,
the accretion luminosity $L_{\rm acc}$ becomes much smaller 
than the stellar luminosity $L_*$. 
Note that the timescale ratio $t_{\rm acc}/t_{\rm KH}$ is 
equal to the luminosity ratio $L_*/L_{\rm acc}$ by definition.
The additional heat input into the atmosphere 
$\eta L_{\rm acc}$ is thus negligible. 


We find that the previous results of HOY12 with the shock boundary 
condition are similar to the current cases with $\eta = 0.1$.
For the shock outer boundary, the specific entropy
of accreting material is determined by solving the flow structure 
across the accretion shock front satisfying jump conditions
\citep[e.g.,][]{SST80}. We have no free parameters in this case.
For the cold-accretion limit with $\eta = 0$, 
the stellar radius initially decreases with increasing stellar mass,
until the star abruptly expands, e.g., 
at $M_* \simeq 50~\msun$ for $\mdot = 1~\msunyr$. 
HOY12 also find similar evolutionary behavior with the photospheric
boundary conditions (\ref{eq:photo_b1}) and (\ref{eq:photo_b2}).
This abrupt expansion in the cold-accretion limit is related
to the extension of a deuterium-burning layer in the stellar interior,
which is studied in detail in \citet{HYO10}.
We shall also examine the interior structure in the context of the 
current results in Sec.~\ref{ssec:intr}.


The lower panel of Figure~\ref{fig:mr_fe} depicts 
the evolution of stellar radius at the lowest accretion rate considered here
$\mdot = 10^{-2}~\msunyr$ with the input energy fractions $\eta = 0.1$ 
and $0.0$ (cases a0.01-e0.1 and a0.01-e0).  
In these cases, the timescale inversion to $t_{\rm KH} < t_{\rm acc}$
occurs around the first local maximum of the radius at $M_* \simeq 15~\msun$.
The star contracts for a while after this point, radiating the energy
away (KH contraction).
As briefly mentioned in Sec.~\ref{ssec:cases}, however, we see a peculiar 
evolution for $M_* \gtrsim 100~\msun$, whereby the stellar radius slowly 
oscillates as the stellar mass increases.
\citet{OP03} show that a similar evolution occurs with
$\mdot \gtrsim 4 \times 10^{-3}~\msunyr$ as the total stellar luminosity
$L_{\rm tot} \equiv L_* + L_{\rm acc}$ approaches the Eddington value
during the KH contraction.  
Whereas \citet{OP03} only show this evolution for
shock outer boundary conditions, we show here that this oscillatory 
behavior occurs even in the extreme case of cold accretion, $\eta = 0$.
Our results suggest that this characteristic evolutionary stage should
emerge only with the moderately high accretion rates
$\mdot \sim 10^{-2}~\msunyr$, regardless of different thermal 
properties of the mass accretion.


The evolution of the interior structure differs significantly
when the mass accretion is extremely cold.
Figure~\ref{fig:intr_1em0En0} shows the evolution 
of the stellar interior structure for the extreme case
of an input energy fraction $\eta = 0$ but with 
the same accretion rate (case a1-e0). 
The evolution of stellar structure deviates strongly 
from the case shown
in Figure~\ref{fig:intr_1em0En0.1}, particularly in the
early stages for $M_* \lesssim 100~\msun$.
Cold mass accretion leads to a much lower average
entropy in the stellar interior, which results in
a smaller stellar radius for $M_* \lesssim 30~\msun$.
The central temperature is higher with the smaller radius
for a given stellar mass, as shown in
Figure~\ref{fig:int_var}-(a); the
central temperature exceeds $10^6$~K before the
stellar mass reaches $10~\msun$.
Deuterium burning begins much earlier than in the case with
$\eta = 0.1$, even before the timescale inversion
due to decreasing opacity in the stellar interior
(see Sec.~\ref{ssec:outline} above).
Figure~\ref{fig:intr_1em0En0.1} also shows that the protostar
is fully convective in this early stage, as the energy 
produced in the nearly opaque interior is transported 
mostly via convection.

Convective mixing transports the recently accreted deuterium
inward to the stellar center, where nuclear fusion
takes place. Central deuterium burning ceases 
soon after a radiative layer emerges in the stellar interior,
which prevents convective mixing 
\citep[this is a so-called radiative barrier, e.g.,][]{PS92}.
Deuterium burning resumes at the bottom of the surface 
convective layer, when the local temperature exceeds 
$1.5 \times 10^6$~K. Deuterium shell burning
injects heat near the surface, which causes the surface 
layer to abruptly inflate at $M_* \simeq 50~\msun$\footnote{
Similar behavior was found in our previous
work \citet{HYO10}, when we followed the evolution of
present-day massive ($M_* \gtrsim 10~\msun$) protostars
with cold mass accretion.}.
The later evolution for $M_* \gtrsim 10^3~\msun$, in particular,
looks similar to the fiducial case shown in 
Figure~\ref{fig:intr_1em0En0.1}.
The evolution of accreting supermassive stars is thus insensitive
to potential variations of the thermal efficiency of mass
accretion, which is modeled with varying $\eta$ in this paper.

\subsubsection{Evolution in the HR Diagram}
\label{ssec:HR}

\begin{figure}
  \begin{center}
\epsscale{1.0}
\plotone{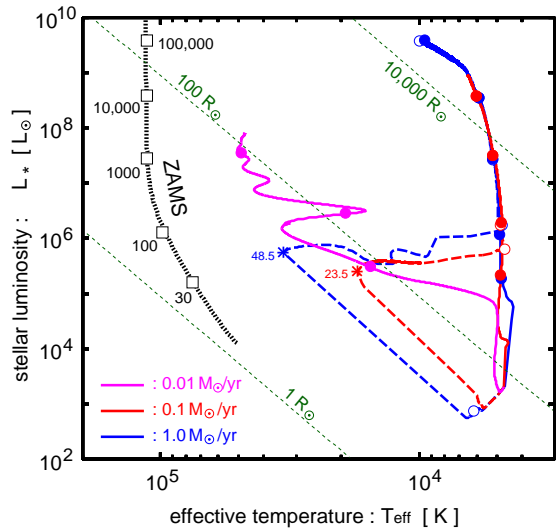}
\caption{Evolutionary tracks in the HR diagram. 
The different colors represent the accretion rates
$1.0~\msunyr$ (blue), $0.1~\msunyr$ (red), and 
$0.01~\msunyr$ (magenta). 
Only the cases with $\eta = 0.1$ 
(cases a1-e0.1, a0.1-e0.1, and a0.01-e0.1) 
and $\eta = 0$ (cases a1-e0 and a0.1-e0) are presented.
The asterisks on the dashed lines mark the points where the
effective temperature takes the highest value for $\eta = 0$.
The stellar masses at these points are also labeled. 
The black dotted line represents the loci of non-accreting zero metallicity
ZAMS stars. The circles and squares on the tracks represent 
positions of $30~\msun$, $100~\msun$, $1000~\msun$,
$10^4~\msun$, and $10^5~\msun$ stars in ascending order. 
Stellar radii at constant values $10^4~\rsun$, 
$100~\rsun$, and $1~\rsun$ are shown by thin green dashed lines.}
\label{fig:HR}
  \end{center}
\end{figure}


Figure~\ref{fig:HR} shows the evolutionary tracks in the 
Hertzsprung-Russell diagram for several representative cases. 
As shown in HOY12 for cases with rapid mass accretion
$\mdot \geq 0.1~\msun$, the tracks are almost vertical
as the stellar mass, luminosity and radius increase for 
$M_* \gtrsim 100~\msun$. 
The tracks for $\mdot = 1.0~\msun$ show a gradual increase
of the effective temperature for $M_* \gtrsim 10^4~\msun$,
finally reaching $\simeq 10^4$~K 
at $M_* = 10^5~\msun$ (see Figure~\ref{fig:mr_tscl}).
Only during the early phases
before the stellar mass (luminosity) exceeds $100~\msun$ 
($10^6~\lsun$), does the effective temperature depend on the
input energy fraction $\eta$, i.e.
lower $\eta$ results in higher effective temperature.
The evolutionary track for $\mdot = 10^{-2}~\msunyr$ 
lies between the above tracks and the ZAMS line.
Despite the apparent differences among these tracks, the stellar 
luminosities for a given mass are nearly the same in all cases for 
$M_* \gtrsim 1000~\msun$, namely the Eddington value.
It is clear that SMSs with extremely high
mass accretion rates will have low effective temperatures
compared to non-accreting or slowly accreting SMSs.
This fact has a number of consequences on star and BH formation
in the early universe, as discussed 
in Sec.~\ref{sec:discussions} below.

\subsection{Stellar Pulsational Instability}
\label{ssec:pi}

\begin{figure}
  \begin{center}
\epsscale{1.0}
\plotone{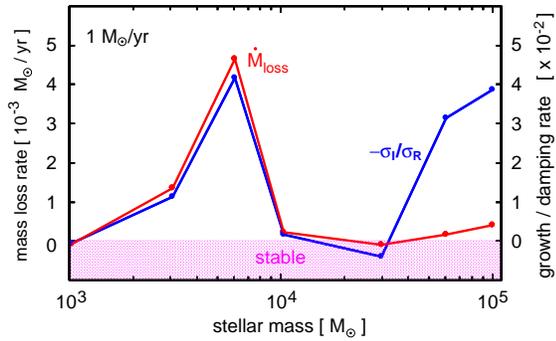}
\caption{
The pulsation stability of an accreting SMS at the rate
$\mdot=1.0~\msunyr$ (case a1-e0.1).
The growth/damping rates of the stellar pulsation
($-\sigma_{\rm I} / \sigma_{\rm R}$) and the resulting
mass-loss rates ($\dot M_{\rm loss}$) are plotted as a 
function of the stellar mass. 
In the shaded region, the star is stable against the stellar pulsation.}
\label{fig:M_Mdot}
 \end{center}
\end{figure}
\begin{figure}
  \begin{center}
\epsscale{1.0}
\plotone{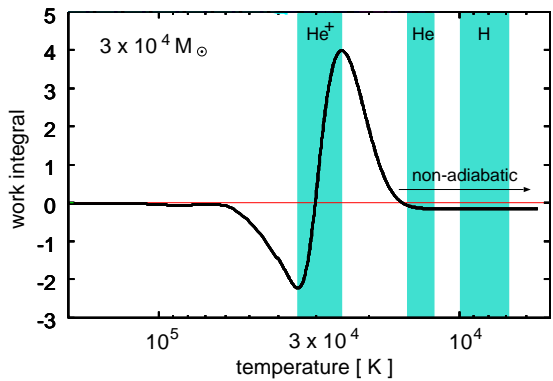}
\caption{The work integral (in arbitrary units, see text) for 
a $3 \times 10^4~\msun$ star accreting at the rate
$\mdot=1.0~\msunyr$ (case a1-e0.1).
Only the distribution near the stellar surface is presented
as a function of the interior temperature. 
The shaded backgrounds represent the ionization layers of He$^+$, 
He, and H. The outermost part is the non-adiabatic layer, 
where the pulsation instability is not excited because 
of radiative diffusion.}
\label{fig:work}
 \end{center}
\end{figure}

Next we study the pulsation stability of accreting SMSs and 
estimate the resulting mass-loss rates for unstable cases.
Following our previous work \citep[IHO13;][]{ISH13},
we apply linear stability analysis to the stellar 
models with different masses.
IHO13 show that an SMS is more unstable 
with higher accretion rates and at higher stellar masses. 
We thus focus on the case with the highest 
accretion rate $\mdot =1.0~\msunyr$ for examining whether the mass loss
caused by the stellar pulsation can limit stellar growth
via mass accretion.
Since accreting SMSs are most unstable against 
radial perturbations for $M_\ast \lesssim 10^3~\msun$ 
(IHO13), we also analyze the stability against radial perturbations.
Our current results allow us to extend the analysis to the
high mass range of $10^3~\msun~\leq  M_\ast  \leq 10^5~\msun$.


The basic procedure of our linear stability analysis is briefly
summarized as follows (see Appendix in IHO13 for more details).
First, we linearize the four {\em time-dependent} stellar structure equations
(i.e., the continuity, momentum, energy, and energy transfer 
equations) and the Poisson equation,
applying radial perturbations to the physical quantities.
We here consider the perturbations proportional to
$e^{i \sigma t}$, where $\sigma =\sigma _{\rm R} + i \sigma _{\rm I}$,
and $\sigma_{\rm R}$ and $|\sigma_{\rm I}|$ are the frequency
and growth (or damping) rate of the pulsation.
For instance, one of the perturbed quantities is the radial 
displacement of a fluid element from the equilibrium position,
$\xi(r,t) \equiv \xi_r(r) e^{i \sigma t}$.
Retaining first order results in a system of ordinary derivative equations
for the perturbed quantities (e.g., $\xi_r(r)$) and $\sigma$.
We solve these equations with appropriate boundary 
conditions, finding solutions for particular eigenvalues of $\sigma$.
The radial profiles of the perturbations are given as
the eigenfunctions of this system of equations.


Figure \ref{fig:M_Mdot} shows the evolution of 
the growth/damping rates of pulsation 
as a function of the accreted mass.
The mass-loss rates are estimated assuming that all the 
pulsation energy is converted into kinetic energy of the outflowing 
gas:
\begin{equation}
\frac{\dot {M}_{\rm loss}}{2}v_{\rm esc}^2=2|\sigma _{\rm I}|E_{\rm W},
\label{eq:massloss}
\end{equation}
where $v_{\rm esc}=(2GM_\ast/R_\ast)^{1/2}$ is the escape velocity 
from the stellar surface, 
\begin{equation}
E_{\rm W}=\frac{\sigma_{\rm R}^2}{2}\int ^{M_\ast}_0 |\xi_r|^2dM_r, 
\end{equation}
is the pulsation energy, 
and $M_r$ the enclosed mass.
Since pulsation energy is partially lost through radiative 
dissipation, our estimate of the mass-loss rate
should be considered an upper limit \citep{Pap73b,Pap73a}.


In agreement with our previous analysis we find that the 
instability is excited by the $\kappa$ mechanism due to the 
opacity bump in the He$^+$ ionization layer.
The accreting SMS is always unstable, except within the narrow mass 
range $10^4~\msun \lesssim M_\ast \lesssim 3 \times 10^4~\msun$.
The mass-loss rate rises with increasing mass
for $M_\ast \lesssim 6 \times 10^3~\msun$. 
The derived mass-loss rate at $M_\ast \sim 10^3~\msun$ is somewhat 
lower than the value shown in IHO13, which we attribute to the 
different numerical code employed in this paper. 
The current stellar model has a slightly higher temperature
in the surface layers than the model analyzed in IOH13.
The layers with the opacity bump, where the pulsation is excited, 
contains a smaller fraction of the stellar mass, which results in
a lower mass-loss rate.
The maximum mass-loss rate is $5\times 10^{-3}~\msunyr$ at 
$M_\ast \simeq 6 \times 10^3~\msun$. 
After this the mass-loss rate hardly rises in spite
of the growth rate increasing with stellar mass.
This is primarily because the escape velocity increases
with stellar mass (see eq.~\ref{eq:massloss}), as the stellar 
expansion slows down for 
$M_* \gtrsim 10^4~\msun$ (Fig.~\ref{fig:mr_tscl}-a).
Since the mass-loss rates are much lower than the assumed accretion rate, 
we conclude that the pulsation instability does not prevent 
the formation of an SMS via rapid mass accretion.


When the stellar mass exceeds $6\times 10^3~\msun$, 
the pulsation is stabilized and the pulsation-driven outflows 
are also suppressed. 
To understand this stabilization, we consider the work integral 
defined by
\begin{equation}
W(M_r)=\frac{\pi}{\sigma _{\rm R}}\int ^{M_r}_0
\Re \left[ 
\frac{\delta T^*}{T}
\left( \delta \epsilon - \frac{d}{dM_r}\delta L_{\rm rad} \right)
\right] dM_r \, ,
\label{eq:work}
\end{equation}
where $\epsilon$ is the nuclear energy production rate per unit mass, 
$L_{\rm rad}$ the radiative luminosity, 
the symbols with $\delta$ represent the Lagrange perturbations, 
and $\Re[~]$ denotes taking the real part of a quantity. 
The work integral physically represents the pulsation energy gained 
during one period within the mass coordinate $M_r$.
The stellar pulsation is excited (damped) in the layer where $dW/dM_r$ 
is positive (negative). 
If the work integral at the stellar surface $W(M_\ast)$ is positive 
(negative), the protostar is unstable (stable).


Figure \ref{fig:work} shows the radial profile of the work integral 
near the surface of a $3 \times 10^4~\msun$ star.
We see that the work integral increases outward within the He$^+$ 
ionization layer, but decreases beyond it because of radiative
diffusion. In the outermost layers with $T \lesssim 18\,000$~K, 
where the radiative cooling time is shorter than the pulsation 
period, the work integral remains constant without excitation 
or damping (so-called the {\it non-adiabatic region}).
In this case, the work integral at the stellar surface 
is negative and the protostar is stable. 
In the most unstable case when $M_\ast= 6 \times 10^3~\msun$,
on the other hand, radiative cooling in the outermost part
is more efficient and the non-adiabatic region extends further inward.
The transition point to the non-adiabatic region is located
just outside of the He$^+$ ionization layer. 
Therefore, the stellar pulsation excited in the He$^+$ ionization 
layer does not suffer from damping via radiative diffusion 
outside the ionization layer. 
Such a profile of the work integral is displayed
in Figure 3 of IHO13, where by
simply extrapolating the available results for $M_\ast < 10^3~\msun$
without considering the damping effect outside the He$^+$ ionization layer,
we had speculated that the mass-loss rates would
increase with stellar mass for $M_\ast > 10^3~\msun$.
By contrast, our current results show that the pulsation
instability is strongly suppressed for $M_\ast \gg 10^3~\msun$ 
and mass-loss is much lower than predicted by IHO13.

Note that the pulsational instability considered above
qualitatively differs from that of non-accreting massive primordial
stars, which is caused by the $\epsilon$-mechanism \citep[e.g.,][]{BHW01,SU12}.
The pulsation driven by the $\epsilon$-mechanism could work when a
SMS contracts and approaches the ZAMS stage after the mass 
accretion ceases (also see Sec.~\ref{ssec:fate}).

\section{Discussion}
\label{sec:discussions}

\subsection{UV feedback from Rapidly Accreting 
Supermassive Stars}
\label{ssec:uvfdbk}

\begin{figure}
  \begin{center}
\epsscale{1.0}
\plotone{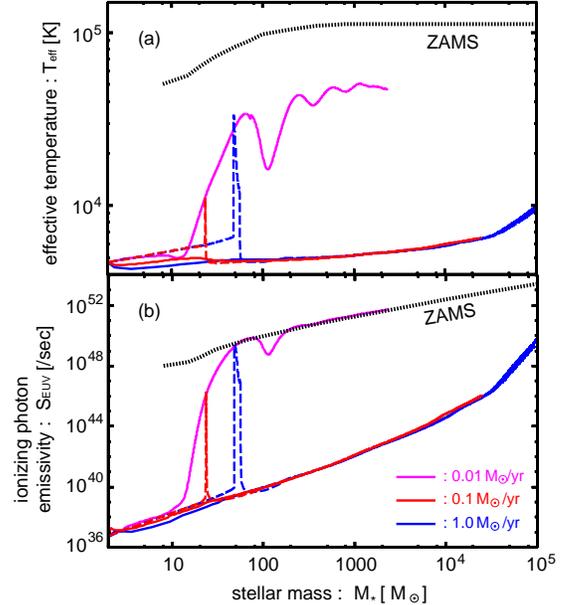}
\caption{Evolution of the stellar effective temperature 
({\it upper panel}) and ionizing photon emissivity 
({\it lower panel}). The lines represent the same cases as 
in Figure~\ref{fig:HR}. The black dotted line presents the 
values of ZAMS stars.}
\label{fig:teff_euv}
  \end{center}
\end{figure}

As mentioned in Sec.~\ref{sec:intro}, for a more typical case of 
primordial star formation, stellar UV feedback becomes 
so strong that it is able to shut off the mass accretion onto the star
\citep[e.g.,][]{HOYY11,HYOY12}.
By contrast, the stellar evolution calculations presented here predict that
this is not the case when forming primordial SMSs via
very rapid mass accretion. 
Figure~\ref{fig:teff_euv} shows that with extremely high accretion 
rates $\mdot \geq 0.1~\msunyr$, the stellar output of 
hydrogen-ionizing photons 
remains very low even after the stellar mass exceeds $10^4~\msun$.
For accretion at $1.0~\msunyr$ the hydrogen-ionizing luminosity for
$M_* \simeq 10^5~\msun$ is still comparable to that of
$M_* \lesssim 100~\msun$ primordial ZAMS stars.
This is a consequence of the low effective temperature resulting
from the large stellar radius.
For $\mdot = 0.1~\msunyr$, the accretion rate of hydrogen atoms 
is $\simeq 3 \times 10^{49}~{\rm sec}^{-1}$.
An HII region cannot grow if the stellar hydrogen-ionizing
photon emissivity is lower than this hydrogen accretion rate.
Only for the extreme cases with low-entropy accretion 
$\eta = 0$, would the ionizing luminosity temporarily increase 
in an early stage when $M_* < 100~\msun$,
visible as a spike in Figure~\ref{fig:teff_euv}. 
This occurs shortly before the abrupt expansion
of the star (Figs.~\ref{fig:mr_fe} and \ref{fig:intr_1em0En0}),
when the stellar total luminosity also sharply rises.
Even if this ``flash'' of ionizing radiation would occur, however,
its duration of $< 100$~years is too short to significantly
disturb the accretion flow.
It is thus unlikely that the mass accretion onto rapidly accreting
SMSs is hindered by stellar UV feedback,
at least for $M_* \lesssim 10^5~\msun$.


After the stellar mass exceeds $10^5~\msun$, however, the stellar 
ionizing flux will continue to increase, so that an HII region might 
finally emerge. The UV feedback caused by the dynamical expansion of 
the HII region would shut off the mass accretion as expected in
the ordinary cases of the primordial star formation 
\citep[e.g.,][]{MT08,HOYY11}. 
The maximum stellar mass by the UV feedback would be higher
with more rapid mass accretion, as suggested in our radiation 
hydrodynamic numerical simulations \citep[][]{HOYY11,HYOY12}.


By contrast, stellar UV feedback at
a lower accretion rate $\mdot = 10^{-2}~\msunyr$ will slow 
and perhaps stop the accretion flow.
Although this rate is too low to form an SMS, 
such moderately rapid accretion can be expected
occasionally in the ordinary mode of primordial star formation
\citep[e.g.,][]{Hirano13}.
Figures~\ref{fig:HR} and \ref{fig:teff_euv}-(a) show that,
in this case, the effective temperature always lies between the ZAMS 
and the rapidly-accreting ($\mdot \geq 0.1~\msunyr$) values.
However, Figure~\ref{fig:teff_euv}-(b) shows that the evolution of the
stellar ionizing luminosity is almost the same as in the ZAMS case.
This is because the larger radius (Fig.~\ref{fig:mr_fe}-b)
compensates for the lower effective temperature,
resulting in a comparable UV luminosity.
Although stellar UV feedback is certainly expected in this case, 
it is somewhat weaker than for the ZAMS stars because of the lower
average energy of ionizing photons.

\subsection{Observational Detectability}

\begin{figure}
  \begin{center}
\epsscale{1.0}
\plotone{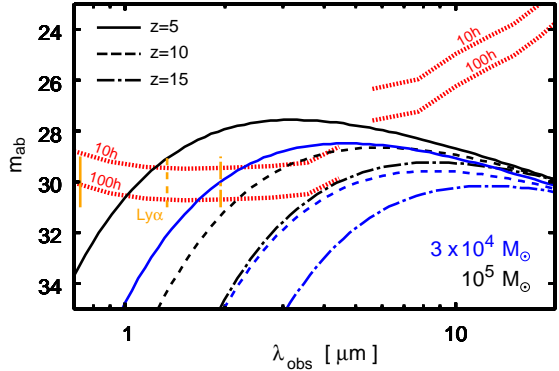}
\caption{Apparent AB magnitudes of bloated supermassive 
stars and detection limits of the James Webb Telescope (JWST).
The black and blue lines represent the $3 \times 10^4~\msun$ and
$10^5~\msun$ stars located at redshifts $z=5$ (solid), 
$10$ (dashed), and $15$ (dot-solid lines). 
The bolometric luminosities of these stars are $1.1 \times 10^9~\lsun$
and $3.8 \times 10^9~\lsun$ respectively.
The red dotted lines show the JWST $10\sigma$ detection limits
with the exposure times of 10 and 100 hours using the 
Near Infrared Camera (NIRCam, left)
and Mid-Infrared Camera (MIRI, right). The JWST photometric 
sensitivity data are taken from the web site, 
http://www.stsci.edu/jwst/science/sensitivity/jwst-phot.
The orange vertical lines mark the wavelength of Ly-$\alpha$ line
at each redshift, indicating that intergalactic absorption 
would reduce the original spectrum for the wavelengths shorter 
than this point. }
\label{fig:jwst}
  \end{center}
\end{figure}

Although the UV luminosity of an accreting SMS is much weaker
than its ZAMS counterpart, the bolometric luminosity 
could exceed $10^9~\lsun$, increasing with 
stellar mass almost linearly.
With such a high luminosity individual SMSs could be
detectable with future observational facilities such as the
James Webb Space Telescope (JWST). Figure~\ref{fig:jwst} demonstrates
this, showing that bloated SMSs at $z \sim 10$ are detectable
by JWST with exposure times of 10 to 100 hours.
The cosmological parameters for the standard $\Lambda$CDM
\citep{Komatsu11} and black-body spectrum with the effective 
temperature $T_{\rm eff}$ are adopted here.


Actually, the observational features of
accreting SMSs are very similar to those of hypothetical
supermassive dark stars (SMDSs), 
which are powered by energy production via annihilation of
dark matter (DM) rather than nuclear fusion 
\citep[e.g.,][]{Spolyar08,Umeda09}.
This is because SMDSs could have a very large radius
$R_* \gtrsim 10$~AU and a low effective temperature
$T_{\rm eff} \lesssim 10^4$~K,
similar to the accreting SMSs presented in this paper.
Recent studies that examined the observational
signatures and detectability of SMDSs \citep[e.g.,][]{Zac10,Freese10,Ilie12}, 
are thus applicable to accreting SMSs as well. 
For instance, SMDS and SMS colors obtained from multi-band survey data 
should be unusual compared to ordinary galaxies and AGNs 
because of their very low effective temperature \citep{Zac10}.
Note that the evolution and apparent colors of SMDSs
are strongly dependent on supply rates of annihilating DM
\citep{Freese10}. 
Once the SMDSs are no longer supplied with sufficient DM, they
contract, increasing their effective temperature.
It is thus only in extreme cases that a $10^5~\msun$ SMDS would still have
a low effective temperature of $T_{\rm eff} \lesssim  10^4$~K.
By contrast, rapid mass accretion keeps SMSs inflated, 
resulting in a low $T_{\rm eff}$ without DM annihilation. 
The brightest sources with unusually red colors might be
a signature of rapidly accreting SMSs rather than of SMDSs.


The detection rate of accreting SMSs depends on their
formation rate per unit redshift and typical lifetime. 
\citet{Agarwal12} and \citet{Johnson13a} show that the formation 
of the SMSs via rapid mass accretion should be more ubiquitous 
than previously thought.
They show that, within the lifetime of $10^6$ yr,  
at least a few accreting SMSs should lie in the survey area of 
$\sim 100~{\rm arcmin}^2$.
The exact lifetime of a SMS depends on the final stage 
of its evolution, which is affected by the mass accretion
history (also see Sec.~\ref{ssec:fate}).

\subsection{The Ultimate Fate of Accreting Supermassive Stars}
\label{ssec:fate}

\begin{figure}
  \begin{center}
\epsscale{1.0}
\plotone{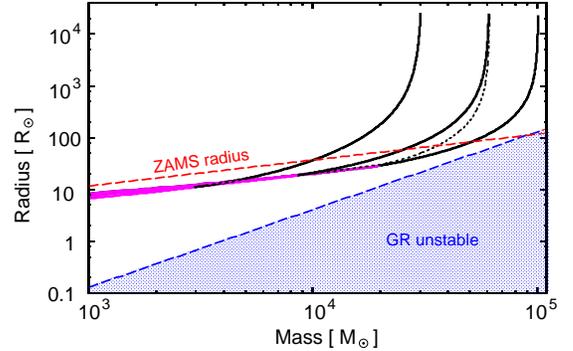}
\caption{
General-relativistic stability of supermassive stars growing
via very rapid mass accretion.
The solid curves depict the radial mass distributions
in the interior of the $3 \times 10^4~\msun$, 
$6 \times 10^4~\msun$, and $10^5~\msun$ stars in case a1-e0.1, 
i.e., with $\mdot = 1.0~\msunyr$ and $\eta = 0.1$.
The dotted curve presents the same profile of
$6 \times 10^4~\msun$ star but for the lower 
accretion rate $0.3~\msunyr$ (case a0.3-e0.1).
The magenta parts of these curves denote the central convective core.
The red dashed line represents the mass-radius relation 
of primordial ZAMS stars \citep[][]{Johnson13c}.
The blue dashed line represents the critical radii of the
GR stability for $n=3$ polytrope stars \citep[][]{Fricke73}, 
below which the stellar structure is unstable.
Note that the horizontal axis denotes the mass coordinate
for the solid and dotted curves and the stellar mass for the dashed lines.
}
\label{fig:greffect}
  \end{center}
\end{figure}

Our results suggest that rapid mass accretion onto SMSs
drastically changes their stellar structure, causing significant
bloating. This could also influence their ultimate fate.
As described in Sec.~\ref{sec:intro}, SMSs exceeding 
$10^5~\msun$ are thought to directly collapse to form massive 
BHs via a GR instability. 
However, previous studies assume that the SMSs are polytropic
stars with index $n=3$, i.e., fully convective stars with 
homogeneous entropy distributions. Our calculations show that this 
is not the case for rapid mass accretion. 
As Figure~\ref{fig:rprof}-(a) shows, in the stellar interior, 
the specific entropy increases outward until reaching the surface 
convective layer. The homogeneous convective core gradually extends 
after hydrogen burning is ignited at the stellar center. 


Whereas GR effects are not taken into account in the 
current calculations, we have also performed a few test runs 
with a gravitational constant $G_{\rm eff}$ modified by 
post-Newton corrections 
\begin{equation}
G_{\rm eff} = G 
\left( 1 + \frac{P}{\rho c^2} + \frac{4 \pi r^3 P}{M c^2}
         + \frac{2 G M}{c^2 r} \right)  
\end{equation}
\citep[e.g.,][]{Fuller86} and see no significant differences from the 
results presented above.
This is demonstrated in Figure~\ref{fig:greffect}, which shows the
stability of SMSs accreting at the rates $\mdot = 1.0~\msunyr$
and $0.3~\msunyr$. The SMSs are still stable as far as we 
have followed their evolution.
Note that, although the entire stellar structure is not well
approximated by a $n=3$ polytrope, the unstable area 
in the figure is applicable to the central convective core.

Extrapolating the current results, we expect that the star becomes 
unstable when the stellar mass reaches a few $\times~10^5~\msun$
for $\mdot = 1.0~\msunyr$.
We expect that not the entire star but only its inner core 
might collapse due to the GR instability. 
This is because the outer part broadly extends with high specific 
entropy obtained with very rapid mass accretion.
If a BH forms as a result of the collapse, an accretion disk should
also form and grow as the surrounding gas falls toward the center.
A quasi-star powered by the energy production
in the accretion disk instead of nuclear fusion
\citep{Begelman08} might finally emerge.
The figure also suggests that the GR instability could work 
for a slightly lower mass assuming a lower accretion rate, because the 
mass distribution in the interior is more centrally condensed
(also see Sec.~\ref{ssec:intr}). 
With significantly lower accretion rates $\mdot \lesssim 0.1~\msunyr$,
however, the star would exhaust its nuclear fuel and collapse directly
to a BH before the GR instability sets in.


Note that the above scenario assumes 
ongoing mass accretion when the GR instability sets in.
As discussed in Sec.~\ref{ssec:uvfdbk}, however, stellar UV feedback 
might halt the mass accretion before the GR instability sets in. 
Without mass accretion, the bloated SMS should contract on
a KH timescale and finally become a ZAMS star, 
which has a homogeneous entropy distribution because of efficient 
convective mixing.
For such $n \simeq 3$ polytrope stars, the GR instability 
should cause most of the stellar matter to collapse to a massive BH 
\citep[e.g.,][]{Fricke73,Fuller86}.
In this case, the stellar lifetime might be somewhat longer 
than in the above case, where the SMS collapses before 
the mass accretion ceases.


Even if a part of the star becomes unstable, 
this does not necessarily mean that a BH forms from the collapse.
It is conceivable that a very energetic supernova explosion results
from runaway nuclear fusion.
Although this is less likely with the lower metallicities 
\citep[e.g.,][]{Montero12}, recent studies suggest that this might 
occur in the early universe \citep[e.g.,][]{Whalen12,Johnson13b}.
Moreover, the evolution of an SMS should also depend on other 
parameters, e.g., the interior rotation and magnetic fields.  
Indeed, \citet{YDL12} show that the ultimate fates of  
$M_* \leq 10^3~\msun$ primordial stars vary with these parameters. 
The dynamics of the final collapse should also depend
on the stellar rotation. \citet{Reisswig13} show that a massive binary 
BH might form through the collapse of a rapidly rotating SMS.
Thus, the final fates of SMSs should be studied further under
consideration of all the potential effects of very rapid mass accretion,
truncation of accretion, runaway nuclear fusion, the GR instability, 
and stellar rotation and magnetic fields.

\subsection{Seeding Supermassive Black Holes with Supermassive Stars}

Besides following the evolution of individual 
accreting SMSs and the resulting formation of massive BHs,
it is also important to consider how often this 
occurs in the early universe. As described in Sec.~\ref{sec:intro}, 
SMSs should form in massive dark halos where H$_2$ molecules 
are destroyed. Previous studies have mostly 
considered that H$_2$ molecules are destroyed by 
photodissociating radiation from neighboring stars, 
showing that the environments for forming 
the SMSs are extremely rare \citep[e.g.,][]{Dijkstra08}. 
Regarding this issue, \citet{Agarwal12} demonstrate that
the occurrence rate of SMS formation should be elevated
when considering the soft UV radiation from Pop II stars, 
and that the number density of massive BHs 
at $z \gtrsim 6$ expected from current observations 
\citep[e.g.,][]{Tr11,Mortlock11} could agree with this scenario.
However, it is still uncertain whether H$_2$ photodissociation 
is the dominant process which causes the SMS formation or not. 
\citet{IO12}, for instance, show that H$_2$ molecules
are easily collisionally dissociated in dense shocks emerging 
in colliding cold accretion flows at the galaxy formation epoch.
This could be more common than the strong photodissociating radiation
field and should be examined in detail in future studies.


Recent numerical simulations are beginning to reveal the dynamical
evolution in an early stage of SMS formation, whereby
a massive circumstellar disk forms as gas falls toward the 
center of a gas cloud \citep[e.g.,][]{RH09,Latif13,Choi13}.
However, several assumptions of the simulations are somewhat 
arbitrary for enabling SMS formation, e.g., 
turning off H$_2$ molecular cooling or elevating the
photodissociating background field by hand.
We need more realistic simulations beginning with the initial
conditions naturally provided by modern cosmology. 


These previous simulations do not show that a SMS with 
$M_* \gg 1000~\msun$ actually forms via rapid mass accretion.
We have shown, for the first time, the formation
of a $M_* \simeq 10^5~\msun$ SMS with the large accretion rates
predicted by previous studies. 
Our results also suggest that stellar radiative
feedback could be effective after the stellar mass exceeds 
$10^5~\msun$, a prediction which needs to be examined with more detailed
simulations. It would be necessary to consistently follow the
growth of the super-giant protostar and the evolution of the accretion flow, 
by including stellar radiative feedback \citep[e.g.,][]{HOYY11}.

\section{Conclusions}
\label{sec:conclusions}

In this paper, we have studied the evolution of rapidly accreting
SMSs, which could be the precursors of SMBHs observed at $z \gtrsim 6$.
To this end, extending our previous work HOY12, we have followed
the evolution until the stellar mass reaches $10^{4-5}~\msun$ by
numerically solving the stellar interior structure taking account of
the effects of the rapid mass accretion. 


In order to avoid numerical convergence difficulties which limited the
calculations for $M_* \lesssim 10^3~\msun$ in HOY12, 
we have employed a different stellar evolution code 
in this paper \citep[][]{YB08}. 
The current calculations agree with our previous results,
showing that a rapidly accreting SMS forms through the 
supergiant-protostar stage.
The stellar radius monotonically increases as the 
stellar mass increases, obeying the analytic mass-radius
relation (\ref{eq:rst_analytic}).
The stellar interior is very inhomogeneous during this stage;
most of the interior contracts radiating the energy away, whereas
a surface layer containing a small fraction of the stellar mass
inflates. The stellar expansion continues even after 
the stellar mass exceeds $10^3~\msun$, until the radius finally 
begins to slightly decrease for $\ga 3 \times 10^4~\msun$. 
The termination of the expansion is inevitable, because
H$^-$ bound-free opacity, which keeps the stellar effective
temperature locked at $\simeq 5000$~K, becomes unavailable
as the density in the stellar surface layers drops below 
$10^{-11}\,{\rm g\, cm}^{-3}$. Nonetheless, the stellar radius 
remains very large, $R_* \simeq 2 \times 10^4~\rsun \simeq 100$~AU 
for $M_* \gtrsim 10^4~\msun$, as long as the rapid mass accretion 
continues.


Our current results suggest that SMSs initially form as very bloated
supergiant stars. 
With this very large radius, the stellar effective temperature
is less than $10^4$~K even after the protostar becomes supermassive.
Stellar UV radiation is much weaker than for non-accreting
ZAMS stars. For instance, the ionizing luminosity of a rapidly accreting 
$10^5~\msun$ star should be less than that of a $100~\msun$ ZAMS star.
Strong UV feedback, which could limit the mass 
accretion onto the star, is thus unlikely to operate in this case.
We have also studied the pulsation stability of accreting 
SMSs with our calculated stellar models. 
Our analyses show that accreting SMSs become pulsation unstable
due to the $\kappa$-mechanism (IHO13), 
but the resulting mass-loss rates are still much lower than the 
accretion rates for $M_* \gtrsim 10^3~\msun$. 
Therefore, we conclude that the SMSs formed via very rapid 
mass accretion are not significantly affected by stellar UV feedback 
or pulsational mass loss.
Once a star becomes supermassive, the stellar bolometric luminosity
could exceed $10^9~\lsun$. Rapidly accreting SMSs are thus in principle
detectable with future observational facilities like JWST. 


The evolution after the birth of the SMSs is critical for
understanding the origins of the SMBHs in the early universe. 
Our calculations show that only the inner core of an accreting SMS 
could become GR unstable as the stellar mass increases.
Extrapolating the current results with the accretion rate of 
$1.0~\msunyr$, for instance, we expect that
the star would suffer from the GR instability when the stellar mass 
reaches a few $\times~10^5~\msun$.
The subsequent evolution is beyond the scope of our current study.
If a BH forms after the inner part of the star gravitationally
collapses, a so-called quasi-star \citep{Begelman08} might emerge. 
However, it is also possible that an energetic supernova explosion
occurs with run-away nuclear fusion during the collapse.
Moreover, as the effective temperature gradually increases,
stellar UV feedback could eventually halt mass
accretion for $M_* \gtrsim 10^5~\msun$, if the emission of hydrogen-ionizing 
photons significantly exceeds the infall rate fo hydrogen atoms. 
If mass accretion stops, the SMS would contract on a KH timescale.
The GR instability could set in as the SMS approaches the ZAMS stage.
The above scenario depends on the late stages of SMS evolution, which
should be investigated further in future studies.

{\acknowledgements 
The authors thank Hideyuki Umeda, Yuichiro Sekiguchi, Neal Turner, 
Dominik Schleicher, and Francesco Palla for fruitful discussions 
and comments. T.H. appreciates the support by Fellowship of the Japan
Society for the Promotion of Science for Research Abroad.
K.I., K.O., and N.Y. are supported by the Grants-in-Aid 
by the Ministry of Education, Science and Culture of Japan 
(23$\cdot$838, 2168407, 21244021, and 25287050).
Portions of this work were conducted at the Jet Propulsion Laboratory,
California Institute of Technology, operating under a contract with 
the National Aeronautics and Space Administration (NASA).}

\bibliography{biblio}

\begin{thebibliography}{80}
\expandafter\ifx\csname natexlab\endcsname\relax\def\natexlab#1{#1}\fi

\bibitem[{{Agarwal} {et~al.}(2012){Agarwal}, {Khochfar}, {Johnson}, {Neistein},
  {Dalla Vecchia}, \& {Livio}}]{Agarwal12}
{Agarwal}, B., {Khochfar}, S., {Johnson}, J.~L., {Neistein}, E., {Dalla
  Vecchia}, C., \& {Livio}, M. 2012, \mnras, 425, 2854

\bibitem[{{Appenzeller} \& {Fricke}(1972{\natexlab{a}})}]{AF72a}
{Appenzeller}, I., \& {Fricke}, K. 1972{\natexlab{a}}, \aap, 18, 10

\bibitem[{{Appenzeller} \& {Fricke}(1972{\natexlab{b}})}]{AF72b}
---. 1972{\natexlab{b}}, \aap, 21, 285

\bibitem[{{Baraffe} {et~al.}(2001){Baraffe}, {Heger}, \& {Woosley}}]{BHW01}
{Baraffe}, I., {Heger}, A., \& {Woosley}, S.~E. 2001, \apj, 550, 890

\bibitem[{{Begelman}(2010)}]{Begelman10}
{Begelman}, M.~C. 2010, \mnras, 402, 673

\bibitem[{{Begelman} {et~al.}(2008){Begelman}, {Rossi}, \&
  {Armitage}}]{Begelman08}
{Begelman}, M.~C., {Rossi}, E.~M., \& {Armitage}, P.~J. 2008, \mnras, 387, 1649

\bibitem[{{Bodenheimer} {et~al.}(2007){Bodenheimer}, {Laughlin},
  {R{\'o}zyczka}, \& {Yorke}}]{Bod07}
{Bodenheimer}, P., {Laughlin}, G.~P., {R{\'o}zyczka}, M., \& {Yorke}, H.~W.,
  eds. 2007, {Numerical Methods in Astrophysics: An Introduction}

\bibitem[{{Bromm} \& {Loeb}(2003)}]{BL03}
{Bromm}, V., \& {Loeb}, A. 2003, \apj, 596, 34

\bibitem[{{Bromm} {et~al.}(2009){Bromm}, {Yoshida}, {Hernquist}, \&
  {McKee}}]{BYHM09}
{Bromm}, V., {Yoshida}, N., {Hernquist}, L., \& {McKee}, C.~F. 2009, \nat, 459,
  49

\bibitem[{{Chandrasekhar}(1964)}]{Chandra64}
{Chandrasekhar}, S. 1964, Physical Review Letters, 12, 114

\bibitem[{{Choi} {et~al.}(2013){Choi}, {Shlosman}, \& {Begelman}}]{Choi13}
{Choi}, J.-H., {Shlosman}, I., \& {Begelman}, M.~C. 2013, \apj, 774, 149

\bibitem[{{Dijkstra} {et~al.}(2008){Dijkstra}, {Haiman}, {Mesinger}, \&
  {Wyithe}}]{Dijkstra08}
{Dijkstra}, M., {Haiman}, Z., {Mesinger}, A., \& {Wyithe}, J.~S.~B. 2008,
  \mnras, 391, 1961

\bibitem[{{Freese} {et~al.}(2010){Freese}, {Ilie}, {Spolyar}, {Valluri}, \&
  {Bodenheimer}}]{Freese10}
{Freese}, K., {Ilie}, C., {Spolyar}, D., {Valluri}, M., \& {Bodenheimer}, P.
  2010, \apj, 716, 1397

\bibitem[{{Fricke}(1973)}]{Fricke73}
{Fricke}, K.~J. 1973, \apj, 183, 941

\bibitem[{{Fuller} {et~al.}(1986){Fuller}, {Woosley}, \& {Weaver}}]{Fuller86}
{Fuller}, G.~M., {Woosley}, S.~E., \& {Weaver}, T.~A. 1986, \apj, 307, 675

\bibitem[{{Haiman}(2013)}]{Haiman13}
{Haiman}, Z. 2013, in Astrophysics and Space Science Library, Vol. 396,
  Astrophysics and Space Science Library, ed. T.~{Wiklind}, B.~{Mobasher}, \&
  V.~{Bromm}, 293

\bibitem[{{Hartmann} {et~al.}(1997){Hartmann}, {Cassen}, \& {Kenyon}}]{HCK97}
{Hartmann}, L., {Cassen}, P., \& {Kenyon}, S.~J. 1997, \apj, 475, 770

\bibitem[{{Hayashi}(1961)}]{Hayashi61}
{Hayashi}, C. 1961, \pasj, 13, 450

\bibitem[{{Henyey} {et~al.}(1964){Henyey}, {Forbes}, \& {Gould}}]{Henyey64}
{Henyey}, L.~G., {Forbes}, J.~E., \& {Gould}, N.~L. 1964, \apj, 139, 306

\bibitem[{{Hirano} {et~al.}(2013){Hirano}, {Hosokawa}, {Yoshida}, {Umeda},
  {Omukai}, {Chiaki}, \& {Yorke}}]{Hirano13}
{Hirano}, S., {Hosokawa}, T., {Yoshida}, N., {Umeda}, H., {Omukai}, K.,
  {Chiaki}, G., \& {Yorke}, H.~W. 2013, ArXiv e-prints:1308.4456

\bibitem[{{Hosokawa} {et~al.}(2011{\natexlab{a}}){Hosokawa}, {Offner}, \&
  {Krumholz}}]{HOK11}
{Hosokawa}, T., {Offner}, S.~S.~R., \& {Krumholz}, M.~R. 2011{\natexlab{a}},
  \apj, 738, 140

\bibitem[{{Hosokawa} \& {Omukai}(2009)}]{HO09}
{Hosokawa}, T., \& {Omukai}, K. 2009, \apj, 691, 823

\bibitem[{{Hosokawa} {et~al.}(2012{\natexlab{a}}){Hosokawa}, {Omukai}, \&
  {Yorke}}]{HOY12}
{Hosokawa}, T., {Omukai}, K., \& {Yorke}, H.~W. 2012{\natexlab{a}}, \apj, 756,
  93

\bibitem[{{Hosokawa} {et~al.}(2011{\natexlab{b}}){Hosokawa}, {Omukai},
  {Yoshida}, \& {Yorke}}]{HOYY11}
{Hosokawa}, T., {Omukai}, K., {Yoshida}, N., \& {Yorke}, H.~W.
  2011{\natexlab{b}}, Science, 334, 1250

\bibitem[{{Hosokawa} {et~al.}(2010){Hosokawa}, {Yorke}, \& {Omukai}}]{HYO10}
{Hosokawa}, T., {Yorke}, H.~W., \& {Omukai}, K. 2010, \apj, 721, 478

\bibitem[{{Hosokawa} {et~al.}(2012{\natexlab{b}}){Hosokawa}, {Yoshida},
  {Omukai}, \& {Yorke}}]{HYOY12}
{Hosokawa}, T., {Yoshida}, N., {Omukai}, K., \& {Yorke}, H.~W.
  2012{\natexlab{b}}, \apjl, 760, L37

\bibitem[{{Iben}(1963)}]{Iben63}
{Iben}, Jr., I. 1963, \apj, 138, 1090

\bibitem[{{Ilie} {et~al.}(2012){Ilie}, {Freese}, {Valluri}, {Iliev}, \&
  {Shapiro}}]{Ilie12}
{Ilie}, C., {Freese}, K., {Valluri}, M., {Iliev}, I.~T., \& {Shapiro}, P.~R.
  2012, \mnras, 422, 2164

\bibitem[{{Inayoshi} {et~al.}(2013{\natexlab{a}}){Inayoshi}, {Hosokawa}, \&
  {Omukai}}]{IHO13}
{Inayoshi}, K., {Hosokawa}, T., \& {Omukai}, K. 2013{\natexlab{a}}, \mnras,
  431, 3036

\bibitem[{{Inayoshi} \& {Omukai}(2011)}]{IO11}
{Inayoshi}, K., \& {Omukai}, K. 2011, \mnras, 416, 2748

\bibitem[{{Inayoshi} \& {Omukai}(2012)}]{IO12}
---. 2012, \mnras, 422, 2539

\bibitem[{{Inayoshi} {et~al.}(2013{\natexlab{b}}){Inayoshi}, {Sugiyama},
  {Hosokawa}, {Motogi}, \& {Tanaka}}]{ISH13}
{Inayoshi}, K., {Sugiyama}, K., {Hosokawa}, T., {Motogi}, K., \& {Tanaka},
  K.~E.~I. 2013{\natexlab{b}}, \apjl, 769, L20

\bibitem[{{Johnson} {et~al.}(2013{\natexlab{a}}){Johnson}, {Dalla}, \&
  {Khochfar}}]{Johnson13a}
{Johnson}, J.~L., {Dalla}, V.~C., \& {Khochfar}, S. 2013{\natexlab{a}}, \mnras,
  428, 1857

\bibitem[{{Johnson} {et~al.}(2013{\natexlab{b}}){Johnson}, {Whalen}, {Even},
  {Fryer}, {Heger}, {Smidt}, \& {Chen}}]{Johnson13b}
{Johnson}, J.~L., {Whalen}, D.~J., {Even}, W., {Fryer}, C.~L., {Heger}, A.,
  {Smidt}, J., \& {Chen}, K.-J. 2013{\natexlab{b}}, \apj, 775, 107

\bibitem[{{Johnson} {et~al.}(2012){Johnson}, {Whalen}, {Fryer}, \&
  {Li}}]{Johnson12}
{Johnson}, J.~L., {Whalen}, D.~J., {Fryer}, C.~L., \& {Li}, H. 2012, \apj, 750,
  66

\bibitem[{{Johnson} {et~al.}(2013{\natexlab{c}}){Johnson}, {Whalen}, {Li}, \&
  {Holz}}]{Johnson13c}
{Johnson}, J.~L., {Whalen}, D.~J., {Li}, H., \& {Holz}, D.~E.
  2013{\natexlab{c}}, \apj, 771, 116

\bibitem[{{Kippenhahn} \& {Meyer-Hofmeister}(1977)}]{Kip77}
{Kippenhahn}, R., \& {Meyer-Hofmeister}, E. 1977, \aap, 54, 539

\bibitem[{{Kippenhahn} {et~al.}(2013){Kippenhahn}, {Weigert}, \& {Weiss}}]{Kip}
{Kippenhahn}, R., {Weigert}, A., \& {Weiss}, A. 2013, {Stellar Structure and
  Evolution: Astronomy and Astrophysics Library.~ISBN
  978-3-642-30255-8.~Springer-Verlag Berlin Heidelberg, 2013}

\bibitem[{{Komatsu} {et~al.}(2011){Komatsu}, {Smith}, {Dunkley}, {Bennett},
  {Gold}, {Hinshaw}, {Jarosik}, {Larson}, {Nolta}, {Page}, {Spergel},
  {Halpern}, {Hill}, {Kogut}, {Limon}, {Meyer}, {Odegard}, {Tucker}, {Weiland},
  {Wollack}, \& {Wright}}]{Komatsu11}
{Komatsu}, E., {et~al.} 2011, \apjs, 192, 18

\bibitem[{{Latif} {et~al.}(2013{\natexlab{a}}){Latif}, {Schleicher}, {Schmidt},
  \& {Niemeyer}}]{Latif13}
{Latif}, M.~A., {Schleicher}, D.~R.~G., {Schmidt}, W., \& {Niemeyer}, J.
  2013{\natexlab{a}}, \mnras, 433, 1607

\bibitem[{{Latif} {et~al.}(2013{\natexlab{b}}){Latif}, {Schleicher}, {Schmidt},
  \& {Niemeyer}}]{Latif13b}
{Latif}, M.~A., {Schleicher}, D.~R.~G., {Schmidt}, W., \& {Niemeyer}, J.~C.
  2013{\natexlab{b}}, ArXiv e-prints:1309.1097

\bibitem[{{Mayer} \& {Duschl}(2005)}]{MD05}
{Mayer}, M., \& {Duschl}, W.~J. 2005, \mnras, 358, 614

\bibitem[{{McKee} \& {Tan}(2008)}]{MT08}
{McKee}, C.~F., \& {Tan}, J.~C. 2008, \apj, 681, 771

\bibitem[{{Montero} {et~al.}(2012){Montero}, {Janka}, \&
  {M{\"u}ller}}]{Montero12}
{Montero}, P.~J., {Janka}, H.-T., \& {M{\"u}ller}, E. 2012, \apj, 749, 37

\bibitem[{{Mortlock} {et~al.}(2011){Mortlock}, {Warren}, {Venemans}, {Patel},
  {Hewett}, {McMahon}, {Simpson}, {Theuns}, {Gonz{\'a}les-Solares}, {Adamson},
  {Dye}, {Hambly}, {Hirst}, {Irwin}, {Kuiper}, {Lawrence}, \&
  {R{\"o}ttgering}}]{Mortlock11}
{Mortlock}, D.~J., {et~al.} 2011, \nat, 474, 616

\bibitem[{{Oh} \& {Haiman}(2002)}]{OH02}
{Oh}, S.~P., \& {Haiman}, Z. 2002, \apj, 569, 558

\bibitem[{{Omukai}(2001)}]{Omukai01}
{Omukai}, K. 2001, \apj, 546, 635

\bibitem[{{Omukai} \& {Palla}(2001)}]{OP01}
{Omukai}, K., \& {Palla}, F. 2001, \apjl, 561, L55

\bibitem[{{Omukai} \& {Palla}(2003)}]{OP03}
---. 2003, \apj, 589, 677

\bibitem[{{Omukai} {et~al.}(2008){Omukai}, {Schneider}, \& {Haiman}}]{OSH08}
{Omukai}, K., {Schneider}, R., \& {Haiman}, Z. 2008, \apj, 686, 801

\bibitem[{{Osaki}(1966)}]{Osaki66}
{Osaki}, Y. 1966, \pasj, 18, 384

\bibitem[{{Palla} \& {Stahler}(1992)}]{PS92}
{Palla}, F., \& {Stahler}, S.~W. 1992, \apj, 392, 667

\bibitem[{{Papaloizou}(1973{\natexlab{a}})}]{Pap73b}
{Papaloizou}, J.~C.~B. 1973{\natexlab{a}}, \mnras, 162, 143

\bibitem[{{Papaloizou}(1973{\natexlab{b}})}]{Pap73a}
---. 1973{\natexlab{b}}, \mnras, 162, 169

\bibitem[{{Regan} \& {Haehnelt}(2009)}]{RH09}
{Regan}, J.~A., \& {Haehnelt}, M.~G. 2009, \mnras, 393, 858

\bibitem[{{Reisswig} {et~al.}(2013){Reisswig}, {Ott}, {Abdikamalov}, {Haas},
  {Moesta}, \& {Schnetter}}]{Reisswig13}
{Reisswig}, C., {Ott}, C.~D., {Abdikamalov}, E., {Haas}, R., {Moesta}, P., \&
  {Schnetter}, E. 2013, ArXiv e-prints:1304.7787

\bibitem[{{Schleicher} {et~al.}(2013){Schleicher}, {Palla}, {Ferrara}, {Galli},
  \& {Latif}}]{Sch13}
{Schleicher}, D.~R.~G., {Palla}, F., {Ferrara}, A., {Galli}, D., \& {Latif}, M.
  2013, ArXiv e-prints:1305.5923

\bibitem[{{Shang} {et~al.}(2010){Shang}, {Bryan}, \& {Haiman}}]{Shang10}
{Shang}, C., {Bryan}, G.~L., \& {Haiman}, Z. 2010, \mnras, 402, 1249

\bibitem[{{Shibata} \& {Shapiro}(2002)}]{Shibata02}
{Shibata}, M., \& {Shapiro}, S.~L. 2002, \apjl, 572, L39

\bibitem[{{Siess} \& {Forestini}(1996)}]{Siess96}
{Siess}, L., \& {Forestini}, M. 1996, \aap, 308, 472

\bibitem[{{Siess} {et~al.}(1997){Siess}, {Forestini}, \& {Bertout}}]{Siess97}
{Siess}, L., {Forestini}, M., \& {Bertout}, C. 1997, \aap, 326, 1001

\bibitem[{{Sonoi} \& {Umeda}(2012)}]{SU12}
{Sonoi}, T., \& {Umeda}, H. 2012, \mnras, 421, L34

\bibitem[{{Spolyar} {et~al.}(2008){Spolyar}, {Freese}, \&
  {Gondolo}}]{Spolyar08}
{Spolyar}, D., {Freese}, K., \& {Gondolo}, P. 2008, Physical Review Letters,
  100, 051101

\bibitem[{{Stacy} {et~al.}(2010){Stacy}, {Greif}, \& {Bromm}}]{Stacy10}
{Stacy}, A., {Greif}, T.~H., \& {Bromm}, V. 2010, \mnras, 403, 45

\bibitem[{{Stacy} {et~al.}(2012){Stacy}, {Greif}, \& {Bromm}}]{Stacy12}
---. 2012, \mnras, 2508

\bibitem[{{Stahler} {et~al.}(1986){Stahler}, {Palla}, \& {Salpeter}}]{SPS86}
{Stahler}, S.~W., {Palla}, F., \& {Salpeter}, E.~E. 1986, \apj, 302, 590

\bibitem[{{Stahler} {et~al.}(1980){Stahler}, {Shu}, \& {Taam}}]{SST80}
{Stahler}, S.~W., {Shu}, F.~H., \& {Taam}, R.~E. 1980, \apj, 241, 637

\bibitem[{{Susa}(2013)}]{Susa13}
{Susa}, H. 2013, \apj, 773, 185

\bibitem[{{Treister} {et~al.}(2011){Treister}, {Schawinski}, {Volonteri},
  {Natarajan}, \& {Gawiser}}]{Tr11}
{Treister}, E., {Schawinski}, K., {Volonteri}, M., {Natarajan}, P., \&
  {Gawiser}, E. 2011, \nat, 474, 356

\bibitem[{{Umeda} {et~al.}(2009){Umeda}, {Yoshida}, {Nomoto}, {Tsuruta},
  {Sasaki}, \& {Ohkubo}}]{Umeda09}
{Umeda}, H., {Yoshida}, N., {Nomoto}, K., {Tsuruta}, S., {Sasaki}, M., \&
  {Ohkubo}, T. 2009, J. Cosmol. Astropart. Phys., 8, 24

\bibitem[{{Unno}(1971)}]{Unno71}
{Unno}, W. 1971, \pasj, 23, 123

\bibitem[{{Whalen} {et~al.}(2012){Whalen}, {Heger}, {Chen}, {Even}, {Fryer},
  {Stiavelli}, {Xu}, \& {Joggerst}}]{Whalen12}
{Whalen}, D.~J., {Heger}, A., {Chen}, K.-J., {Even}, W., {Fryer}, C.~L.,
  {Stiavelli}, M., {Xu}, H., \& {Joggerst}, C.~C. 2012, ArXiv
  e-prints:1211.1815

\bibitem[{{Whalen} {et~al.}(2013{\natexlab{a}}){Whalen}, {Johnson}, {Smidt},
  {Heger}, {Even}, \& {Fryer}}]{Whalen13b}
{Whalen}, D.~J., {Johnson}, J.~L., {Smidt}, J., {Heger}, A., {Even}, W., \&
  {Fryer}, C.~L. 2013{\natexlab{a}}, ArXiv e-prints:1308.3278

\bibitem[{{Whalen} {et~al.}(2013{\natexlab{b}}){Whalen}, {Johnson}, {Smidt},
  {Meiksin}, {Heger}, {Even}, \& {Fryer}}]{Whalen13}
{Whalen}, D.~J., {Johnson}, J.~L., {Smidt}, J., {Meiksin}, A., {Heger}, A.,
  {Even}, W., \& {Fryer}, C.~L. 2013{\natexlab{b}}, \apj, 774, 64

\bibitem[{{Wise} {et~al.}(2008){Wise}, {Turk}, \& {Abel}}]{WTA08}
{Wise}, J.~H., {Turk}, M.~J., \& {Abel}, T. 2008, \apj, 682, 745

\bibitem[{{Wolcott-Green} {et~al.}(2011){Wolcott-Green}, {Haiman}, \&
  {Bryan}}]{WH11}
{Wolcott-Green}, J., {Haiman}, Z., \& {Bryan}, G.~L. 2011, \mnras, 418, 838

\bibitem[{{Yoon} {et~al.}(2012){Yoon}, {Dierks}, \& {Langer}}]{YDL12}
{Yoon}, S.-C., {Dierks}, A., \& {Langer}, N. 2012, \aap, 542, A113

\bibitem[{{Yorke} \& {Bodenheimer}(2008)}]{YB08}
{Yorke}, H.~W., \& {Bodenheimer}, P. 2008, in Astronomical Society of the
  Pacific Conference Series, Vol. 387, Massive Star Formation: Observations
  Confront Theory, ed. H.~{Beuther}, H.~{Linz}, \& T.~{Henning}, 189

\bibitem[{{Zackrisson} {et~al.}(2010){Zackrisson}, {Scott}, {Rydberg}, {Iocco},
  {Edvardsson}, {{\"O}stlin}, {Sivertsson}, {Zitrin}, {Broadhurst}, \&
  {Gondolo}}]{Zac10}
{Zackrisson}, E., {et~al.} 2010, \apj, 717, 257

\bibitem[{{Zinnecker} \& {Yorke}(2007)}]{ZY07}
{Zinnecker}, H., \& {Yorke}, H.~W. 2007, \araa, 45, 481

\end{thebibliography}
\bibliographystyle{apj}

\end{document}